\documentclass{aastex63}
\usepackage{multirow,booktabs}
\usepackage[utf8]{inputenc}

\shorttitle{An X-ray View of Two IRDCs}
\shortauthors{Yu et al.}

\begin{document}

\title{An X-ray View of Two Infrared Dark Clouds G034.43+00.24 and G035.39-00.33}

\author{Hanbo Yu}
\affiliation{Department of Astronomy, Xiamen University, Xiamen, 361005, China}

\author{Junfeng Wang}
\affiliation{Department of Astronomy, Xiamen University, Xiamen, 361005, China}

\author{Jonathan C. Tan}
\affiliation{Department of Space, Earth \& Environment, Chalmers University of Technology, SE-412 96 Gothenburg, Sweden}
\affiliation{Department of Astronomy, University of Virginia, Charlottesville, VA 22904, USA}

\correspondingauthor{J. Wang}
\email{jfwang@xmu.edu.cn}

\def\chandra{\emph{Chandra}}
\def\spitzer{\emph{Spitzer}}

\begin{abstract}
We present a high spatial resolution \chandra{} X-ray study of two Infrared Dark Clouds (IRDCs), G034.43+00.24 and G035.39-00.33, which are expected to be in the early phases of star cluster formation. We detect 112 and 209 valid X-ray point sources towards G034.43+00.24 and G035.39-00.33, respectively.  We cross-match the X-ray point sources with 2MASS, GLIMPSE and WISE catalogs and find 53\% and 59\% of the X-ray sources in G034.43+00.24 and in G035.39-00.33 have corresponding infrared counterparts, respectively.  These sources are probable members of young massive clusters in formation, and using stellar isochrones we estimate that a population of 1-2 Myr old, intermediate to high mass young stellar objects (YSOs) exist in both IRDCs.  Two and ten Class \uppercase\expandafter{\romannumeral2} counterparts to X-ray sources were identified in G034.43+00.24 and in G035.39-00.33, respectively, which are located in or near dark filaments.  The X-ray Luminosity Function (XLF) of G035.39-00.33 implies that the total mass consists of up to $\sim1,700\:M_\odot$ of stars, using the XLF of the well-studied Orion Nebula Cluster as calibrator. This corresponds to a star formation efficiency of at most $5\%$, indicating the system is still very much gas dominated and in an early stage of the star formation process. The population of G034.43+00.24 is less well determined due to the lower sensitivity of its observations. 
\end{abstract}


\section{Introduction} \label{sec:intro}

Infrared Dark Clouds (IRDCs) are cold, dense molecular gas clumps that are typically observed
near the Galactic plane, i.e., as shadows against the bright Galactic
background. They were discovered in the 1990s as a new, unexpected
component of the cold interstellar medium (ISM)
\citep{perault96,egan98}, and large numbers have now been cataloged
\citep{simon06a,peretto09}.

IRDCs are thought to likely represent the early stages of star cluster
formation given their similar masses and densities as more evolved
cluster-forming clumps \citep[e.g.,][]{rathborne06,henshaw14} in giant molecular clouds (GMCs).
They typically exhibit filamentary structures over a wide range of
scales, including sub-filaments \citep{henshaw16}. Relatively strong
magnetic fields have been inferred to be present in a few IRDCs, 
with field strengths estimated to be $\sim 100 - 1000\:{\rm \mu G}$ \citep[e.g.,][]{pillai15}.
The dynamical evolution of IRDCs is likely to be influenced by a
combination of turbulence, large-scale flows, magnetic field support
and self-gravity \citep[e.g.,][]{henshaw16}.
Some studies pointed out that the filaments of IRDCs have not reached virial 
equilibrium, and the velocity gradients of IRDCs can be considered as a function
of size scale \citep{hernandez11,hernandez15}.  Some relevant theoretical work on IRDCs includes the
magneto-hydrodynamical (MHD) simulation of \citet{li2015}, which was
designed to form a dense, filamentary cloud. The cloud subsequently formed a population of young stellar objects (YSOs). \citet{wu17,wu20} have presented
simulations of magnetised GMC-GMC collisions, which also lead to formation of dense filaments that spawn YSO populations.

G034.43+00.24 \citep[hereafter G34.4; also known as Cloud F in the sample of][]{butler2009} 
is an extensively studied filamentary IRDC. G34.4
was first identified by \citet{miralles1994} in $\rm NH_3$ emission,
present as an elongated structure located to the north of the bright IRAS
source 1807+0121. This structure was then identified as an IRDC in MSX
images \citep{simon06a}. While a parallax distance of 1.56 kpc based on masers has been claimed \citep{kurayama}, the kinematic distance is much larger at 3.7~kpc \citep{forster,xu16}. Given this discrepancy, the distance to G34.4 has been discussed by \citet{forster2012}, who derived an extinction based distance that is consistent with the kinematic distance for this cloud. They commented that the uncertainty of the maser parallax distance may be underestimated. Thus in our work we adopt $d= 3.7$~kpc as the preferred value. With this distance the total mass of the cloud from MIR and NIR extinction mapping is $\sim4,800\:M_\odot$
\citep{kainulainen}. The kinematic structures of G34.4 have been
studied with dense gas tracers, including detection of multiple
velocity components \citep{jonathan18}.

G035.39-00.33 \citep[hereafter G35.4; also known as Cloud H in the sample of][]{butler2009}
 is a relatively massive \citep[$\sim 16,700\:M_\odot$,][]{kainulainen} IRDC
in the W48 molecular cloud, located 2.9 kpc away \citep{simon06b}. It is also noted in the literature to be highly
filamentary, like Cloud F.
There have been numerous studies of the kinematics, dynamics and
chemistry of Cloud H \citep[e.g.,][]{henshaw14,jonathan17}.

Our goal in this paper is to probe the YSO content of the IRDCs, in
particular using {\it Chandra} X-ray observations. YSOs have
traditionally been identified by the presence of infrared (IR) excess \citep{lada87,wolk96}.
However, slightly more evolved YSOs, i.e., Class \uppercase\expandafter{\romannumeral3} sources, do not show
significant excess in the IR and so can be difficult to identify, especially in
crowded regions, such as near the Galactic plane.  X-ray observations
have been shown to be efficient in detecting such diskless stars, as
well as finding a significant fraction of Class
\uppercase\expandafter{\romannumeral1} and
\uppercase\expandafter{\romannumeral2} sources
\citep{feigelson1999,feigelson13}. The contamination from older field
stars in X-ray images of star-forming regions is drastically smaller
than those in optical and IR images, because the X-ray emission of old
Galactic stars is $\sim {10}^{2}$ to $10^3$ times smaller than those of pre-main-sequence (PMS) stars \citep[][]{preibisch05}.

Overall if the number of diskless YSOs (Class \uppercase\expandafter{\romannumeral3}) relative to YSOs with circumstellar disks is low in a star cluster, it likely indicates that such a system is young (less than 1~Myr) and vice versa \citep{H01,jfw07,jfw08}.
As revealed from \chandra{} imaging of some distant star-forming
regions, young star clusters, including massive stars, are often best
revealed by deep X-ray observations \citep{feigelson2007,broos13,povich13}. The
contamination from background quasars is generally much less than $10\%$.  X-ray
flaring activity occurs in all YSO classes \citep{feigelson1999}, and a flux limited X-ray observation to measure an X-ray luminosity
function (XLF) can be related to a YSO population down to a
corresponding mass and optical/NIR magnitude 
\citep[see for example, Massive Young Star Forming Complex in Infrared and X-ray (MYStIX) survey;][]{feigelson13,kuhn2013}. 
It was suggested that the shape of the stellar XLF in young star clusters appears to be universal, which could be further examined to test the uniformity of the stellar initial mass function (IMF) \citep{feig05}. Previous studies find that stellar mass is by far the dominant factor of PMS X-ray luminosity \citep{flaccomio2003,stassun2004,Gudel07}, and such universality relies on the saturation of X-ray luminosity at a fraction of 
the bolometric luminosity in the PMS stars \citep[$\log L_X/L_{bol}\approx -3$;][]{feigelson2002,flaccomio2003}.

The first IRDC observed by {\it Chandra} was G014.225-00.506
\citep{povich10}. A joint analysis of IR and X-ray data was made later
\citep{povich16}.  The target includes an ultracompact
H\uppercase\expandafter{\romannumeral2} region located in the midst of
the M17 SWex IRDC lanes.  The study combined a 98.79 ks ACIS exposure
with mid-infrared data form GLIMPSE \citep{glimpse1,glimpse2} and MIPSGAL \citep{mipsgal}.
By identifying IR excess, which is produced by circumstellar
disks and/or infalling envelopes, and crossmatching to the X-ray
source catalog, a more reliable YSO census was made. \citet{povich16}
 concluded that the youngest and most massive YSOs are
forming in the filamentary structures of the IRDC, while slightly
older diskless intermediate-mass YSOs are distributed in the
surrounding lower-density regions.

Here we present an investigation to assess the pre-main-sequence
population of G34.4 and G35.4 and their immediate environs based on
our X-ray data (PI: J.~Tan). The \chandra{} observations and analysis methods are
described in Section 2.  The association between X-ray sources and
near-infrared data, including the infrared colors of X-ray
counterparts, are presented in Section 3.  The overall cluster
populations inferred from the X-ray luminosity function are discussed
in Section 4. We summarize our findings in Section 5.

\section{\chandra{} Observations and Data Reduction} \label{sec:chandra}

\begin{figure*}
\gridline{\fig{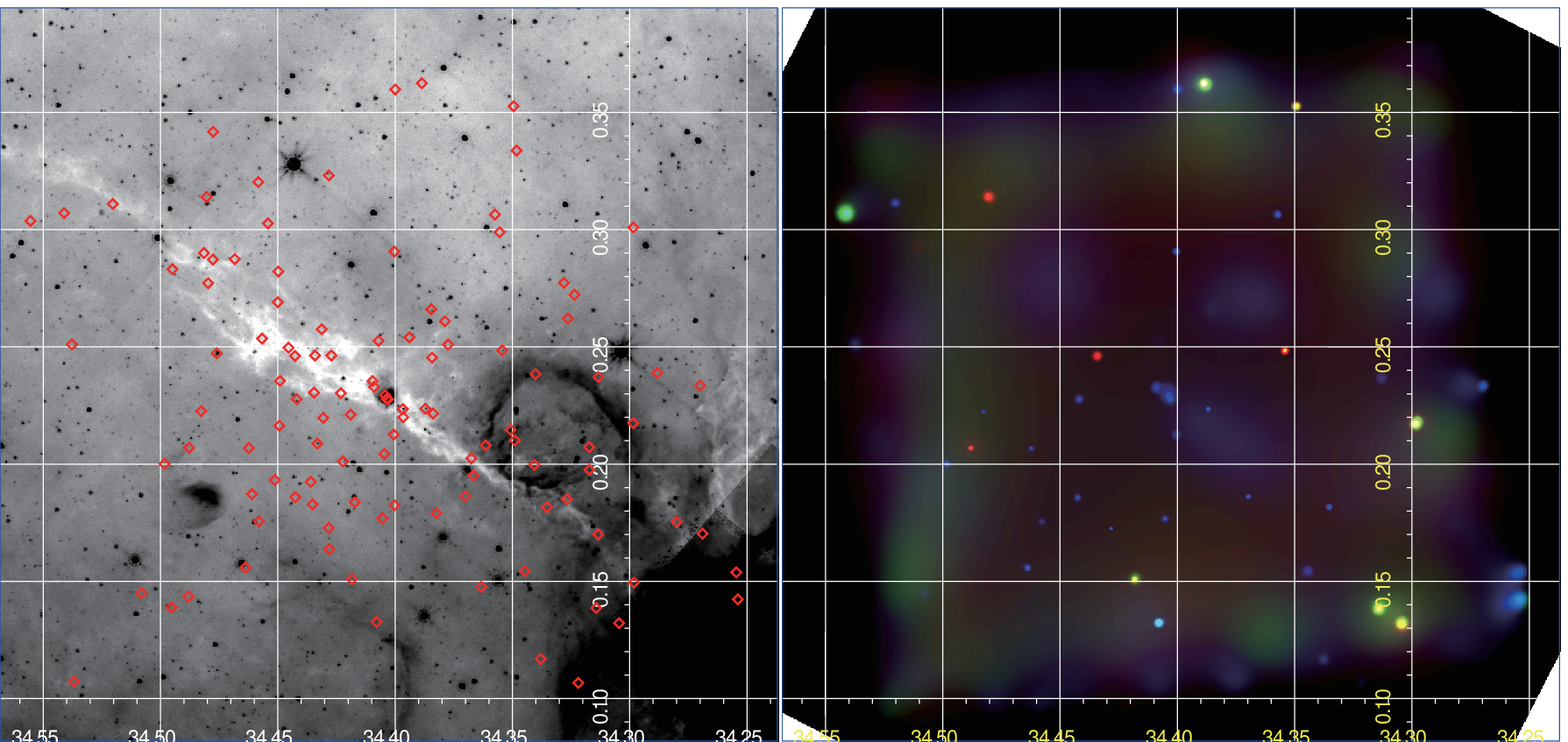}{\textwidth}{(a)}}
\gridline{\fig{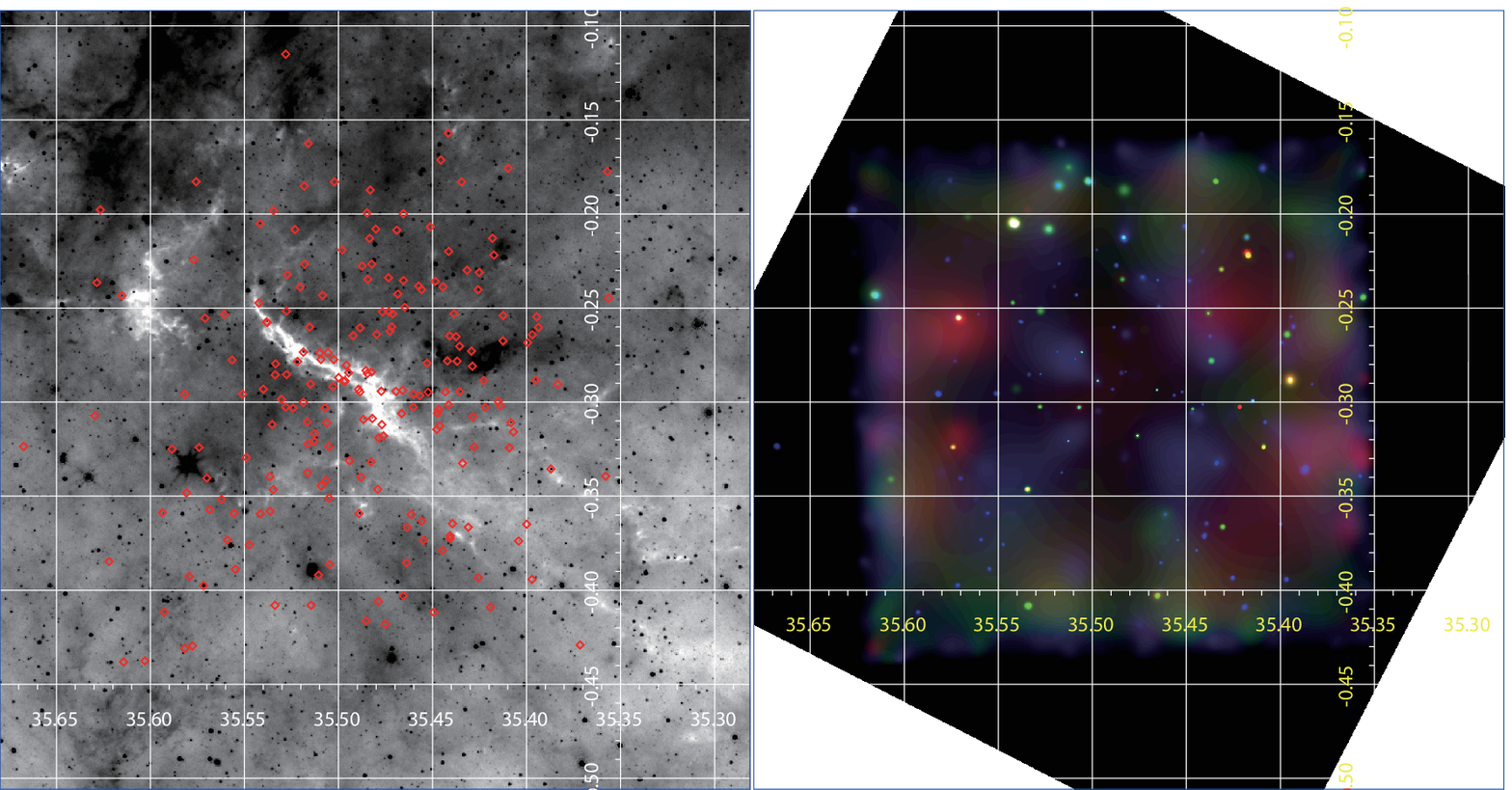}{\textwidth}{(b)}}
\caption{(a) Inverted grey scale \spitzer{} $8.0\mu m$ image (left) and full field \chandra{} images (right) of IRDC G34.4 with a $\sim \ 20^{\prime}\times 20^{\prime}$ field of view. The \chandra{} ACIS-I images were adaptively smoothed by CIAO tool  $csmooth$ \citep{ebeling06}, with parameters \emph{sigmin}=3 and \emph{sigmax}=5. In the smoothed \chandra{} images, red, green and blue represent the soft band (0.5--1.2 keV), medium band
(1.2--2.0 keV) and the hard band (2.0--7.0 keV) emission, respectively. Overlaid are the positions of valid X-ray sources shown as red diamonds. (b) The same as (a), but for IRDC G35.4.  
\label{fig:chandraimage}}
\end{figure*}

IRDCs G34.4 and G35.4 were observed (PI: Tan) in 2013 and 2017, respectively, with the Imaging Array of the
Advanced CCD Imaging Spectrometer (ACIS-I) on board \chandra{}. The field of view of ACIS-I is $17' \times 17'$ in a single pointing
\citep[\url{http://cxc.harvard.edu/cal/Acis/index.html}, more details of the telescope and instruments can be found in ][]{chandra}.  As the exposure
time and roll angle were different in the different sub-observations of the same target, basic information of these observations is listed in Table 1.
Total exposure time is 63~ks for G34.4 and 182~ks for G35.4.
For all observation IDs
(OBSIDs), all four CCDs on ACIS-I were working in the ``Timed Event (TE)'' mode during entire exposure.  OBSIDs
involved in the source searching processes are shown in Table
\ref{tab:basicinfo}.  We followed the standard data reduction procedures using the Chandra Interactive Analysis of Observations (CIAO) software package version 4.10 and the calibration data base (CalDB) version 4.8.5, provided by the Chandra X-ray Center (CXC).  We reprocessed all the event files using CIAO script $chandra\_repro$ to apply relevant corrections and calibrations. Figure \ref{fig:chandraimage} shows the full-field ACIS-I array image at a reduced resolution of 2 arcsec (binning of 4 instrumental pixels). The images were adaptively smoothed using CIAO tool {\it csmooth} (Ebeling et al. 2006).  We note that neither of the IRDCs show clear signal of shadowing against the diffuse soft X-ray background, as they do in the MIR bands, which is likely because of the lack of sensitivity and presence of structures induced by the CCD chip gaps.

\begin{deluxetable*}{ccccccc}
\tablenum{1}
\tablecaption{Log of \chandra{} Observations of IRDCs G34.4 and G35.4\label{tab:basicinfo}}
\tablewidth{0pt}
\tablehead{
\multirow{2}{*}{TARGET}&\multirow{2}{*}{OBSID}&\multirow{2}{*}{\shortstack{START TIME\\(UT)}} & \multirow{2}{*}{\shortstack{EXPOSURE \\TIME(s)}} & \multicolumn{2}{c}{AIM POINT}& \multirow{2}{*}{\shortstack{ROLL ANGLE\\(deg)}}\\
&&&&${\alpha}_{J2000.0}$&${\delta}_{J2000.0}$&
}
\decimalcolnumbers
\startdata
\multirow{2}{*}{G34.4+0.23 (IRDC F)}&14541&2013 Jun 17 12:03:12&28498&283.32&1.42&146.90\\
                                         &15664&2013 Aug 12 17:44:28&34970&283.33&1.40&239.21\\
\hline
\multirow{5}{*}{G35.4-0.33 (IRDC H)}&18921&2017 Jun 19 06:18:36&56074&284.29&2.13&151.21\\
                                    &20100&2017 Jun 21 14:17:43&43989&284.29&2.13&151.21\\
                                    &20101&2017 Jun 24 12:22:34&35080&284.29&2.13&151.21\\
                                    &20102&2017 Jun 25 18:29:03&17115&284.28&2.13&151.21\\
                                    &20103&2017 Jun 08 01:57:22&30069&284.29&2.12&186.65\\
\enddata
\end{deluxetable*}

\subsection{Source Finding and Photon Event Extraction} \label{sec:sourcefinding}

We follow the same customized X-ray point source finding and source extraction described in \citet{jfw08}.
The event files of different ObsIDs of the same target were merged. Two images with different binning scales were created, one was the events in
the full I-array binned by 4 times the ACIS pixel, and the other was produced from central region with no additional binning. The {\em wavdetect} program
\citep{wavdetect} was run with wavelet scales from 1 to 16 pixels in steps of $\sqrt{2}$ (for the two different binning scales, respectively) and a source significance threshold of $10^{-5}$ on each of the original images and merged images, which should yield an expectation value of 10 spurious sources over the entire field on a statistical basis. Thus with such a threshold, the observation is able to find real sources, but has potential contamination at a level of about 10 false detections per field. Thus candidate sources are further evaluated to help eliminate false detections (see below). Source positions from different images were compared in pairs, and duplicate sources were removed if the distance between two sources was measured to be less than the positional uncertainty according to their off-axis angles.  These source lists were merged to generate a master candidate detection list. 

The source finding process resulted in 161 potential sources from 2 ObsIDs for G34.4 and 353 potential sources from 5 ObsIDs for G35.4.  A preliminary event extraction for these potential X-ray sources was
made with IDL script {\em ACIS Extract}\citep[\url{http://personal.psu.edu/psb6/TARA/AE.html}  hereafter AE, ][]{broos02}.  Source list produced by the procedure described above was provided to $AE$ as input.  Validity of the sources in the two
source lists was evaluated by calculating the probability of each source being a false detection ($P_B$), which was solely in accordance
with the Poisson fluctuation of the background.  Those sources with
$P_B > 0.5\%$ were removed from their lists because of their high
likelihoods of being background fluctuations.  The numbers of such
``valid sources'' of the two IRDCs after this cut was 112 for G34.4
and 209 for G35.4 (source positions shown in \spitzer{} image of Figure \ref{fig:chandraimage}). 
Information of these valid sources is presented in
Table \ref{tab:g344info} and Table \ref{tab:g354info} for the two
IRDCs. 

\startlongtable
\begin{deluxetable*}{cccccccccccccccccc}
\tablecaption{Basic Source Information of G34.4\label{tab:g344info}}
\tablewidth{700pt}
\tablenum{2}
\tabletypesize{\tiny}
\tablehead{\\
\multicolumn{2}{c}{\multirow{2}{*}{Source}}&\multicolumn{4}{c}{\multirow{2}{*}{Position}}&\multicolumn{5}{c}{\multirow{2}{*}{Extracted Counts}}&\multicolumn{7}{c}{\multirow{2}{*}{Characteristics}}\\
    \cmidrule(r){3-6}
    \cmidrule(r){7-11}
    \cmidrule(r){12-18}
    &&
    \multirow{2}{*}{\shortstack{$\alpha_{J2000.0}$\\(deg)}}&\multirow{2}{*}{\shortstack{$\delta_{J2000.0}$\\(deg)}}&\multirow{2}{*}{\shortstack{Err\\('')}}&\multirow{2}{*}{\shortstack{$\theta$\\(')}}&
    \multirow{2}{*}{\shortstack{Net\\Full}}&\multirow{2}{*}{\shortstack{$\Delta$Net\\Full}}&\multirow{2}{*}{\shortstack{Bkgd\\Full}}&\multirow{2}{*}{\shortstack{Net\\Hard}}&\multirow{2}{*}{\shortstack{PSF\\Frac}}&
    \multirow{2}{*}{Signif}&\multirow{2}{*}{log$P_B$}&\multirow{2}{*}{Anom}&\multirow{2}{*}{Var}&\multirow{2}{*}{\shortstack{ Exp\\(ks)}}&\multirow{2}{*}{\shortstack{$E_{m}$\\(keV)}}&\multirow{2}{*}{\shortstack{Hardness\\Ratio}}\\
    \cmidrule(r){1-2}
    Label&CXOU J\\
    (1)&(2)&(3)&(4)&(5)&(6)&(7)&(8)&(9)&(10)&(11)&(12)&(13)&(14)&(15)&(16)&(17)&(18)
} 
\startdata
1  &185246.09+012520.5&283.192047&1.422383&0.6&7.9&31.2&6.0&124.0&3.2&0.9&4.9&\textless-5&g&a&62.7&1.2&-0.61\\
2  &185248.28+012742.0&283.201172&1.461688&0.3&7.9&90.1&9.9&100.0&25.8&0.9&9.0&\textless-5&&b&62.7&1.5&-0.36\\
3  &185249.99+012445.6&283.208313&1.412682&0.7&6.9&9.6&3.7&100.0&8.2&0.9&2.4&-4.1&g&a&62.7&5.0&0.69\\
4  &185250.09+012814.0&283.20871&1.470556&0.6&7.7&26.8&5.9&108.0&25.5&0.9&4.3&\textless-5&g&a&62.7&3.7&0.82\\
5  &185251.54+012112.1&283.214783&1.353387&0.7&7.3&15.9&4.5&102.0&12.2&0.9&3.3&\textless-5&g&a&62.7&3.3&0.50\\
6  &185256.83+012430.3&283.236816&1.408417&0.3&5.2&24.7&5.1&101.0&25.1&0.9&4.5&\textless-5&&a&62.7&5.3&1.00\\
7  &185258.22+012411.5&283.242615&1.403208&0.6&4.8&4.0&2.3&103.0&1.3&0.9&1.5&-2.3&g&a&62.7&2.0&-0.20\\
8  &185259.83+012208.2&283.249329&1.36895&0.6&5.0&6.0&2.7&114.0&3.2&0.9&2.0&-4.0&&a&62.7&3.1&0.14\\
9  &185300.43+012145.6&283.251801&1.362685&0.5&5.1&8.7&3.2&132.0&7.0&0.9&2.5&\textless-5&&a&62.7&2.5&0.60\\
10 &185301.02+012844.5&283.254272&1.479045&0.6&5.8&7.7&3.2&315.0&8.3&0.9&2.2&-3.8&&b&62.7&3.4&1.00\\
\enddata
\tablecomments{Basic information of sources in G34.4 is presented here:
Col. (1): X-ray catalog sequence number, sorted by right ascension. 
Col. (2): IAU designation. 
Cols. (3) and (4): Right ascension and declination for epoch J2000.0.
Col. (5): Estimated random component of position error, $1\delta$, computed as (standard deviation of PSF inside extraction region)/(number of counts extracted)$^{1/2}$. 
Col. (6): Offaxis angle. 
Cols. (7) and (8): Estimated net counts extracted in the total energy band (0.5-8 keV ) and average of the upper and lower $1\delta$ errors on col. (7). 
Col. (9): Background counts extracted (total band). 
Col. (10): Estimated net counts extracted in the hard energy band (2-8 keV ). 
Col. (11): Fraction of the PSF (at 1.497 keV ) enclosed within the extraction region. Note that a reduced PSF fraction (significantly below 90\%) may indicate that the source is in a crowded region. 
Col. (12): Photometric significance computed as (net counts)/(upper error on net counts). 
Col. (13): Log probability that extracted counts (total band) are solely from background. Some sources have PB values above the 1\% threshold that defines the catalog because local background estimates can rise during the final extraction iteration after sources are removed from the catalog. 
Col. (14): Source anomalies: g = fractional time that the source was on a detector ( FRACEXPO from mkarf) is <0.9; e = source on field edge; p = source piled up; s = source on readout streak.
Col. (15 ): Variability characterization based on K-S statistic (total band): a = no evidence for variability ($0.05<P_{K-S}$); b = possibly variable ($0.005<P_{K-S}<0.05$); c = definitely variable ($P_{K-S}<0.005$). No value is reported for sources with fewer than 4 counts or for sources in chip gaps or on field edges. 
Col. (16 ): Effective exposure time: approximate time the source would have to be observed on axis to obtain the reported number of counts. 
Col. (17): Background-corrected median photon energy (total band).
Col. (18): Hardness Ratio defined as $\frac{H-S}{H+S}$, where H and S correspond to source counts of hard band and soft band.
Table 3 is available in its entirety in its machine readable format. 
A portion is shown here for guidance regarding its form and content.}
\end{deluxetable*}

\startlongtable
\begin{deluxetable*}{cccccccccccccccccc}
\tablecaption{Basic Source Information of G35.4\label{tab:g354info}}
\tablewidth{700pt}
\tablenum{3}
\tabletypesize{\tiny}
\tablehead{\\
\multicolumn{2}{c}{\multirow{2}{*}{Source}}&\multicolumn{4}{c}{\multirow{2}{*}{Position}}&\multicolumn{5}{c}{\multirow{2}{*}{Extracted Counts}}&\multicolumn{7}{c}{\multirow{2}{*}{Characteristics}}\\
    \cmidrule(r){3-6}
    \cmidrule(r){7-11}
    \cmidrule(r){12-18}
    &&
    \multirow{2}{*}{\shortstack{$\alpha_{J2000.0}$\\(deg)}}&\multirow{2}{*}{\shortstack{$\delta_{J2000.0}$\\(deg)}}&\multirow{2}{*}{\shortstack{Err\\('')}}&\multirow{2}{*}{\shortstack{$\theta$\\(')}}&
    \multirow{2}{*}{\shortstack{Net\\Full}}&\multirow{2}{*}{\shortstack{$\Delta$Net\\Full}}&\multirow{2}{*}{\shortstack{Bkgd\\Full}}&\multirow{2}{*}{\shortstack{Net\\Hard}}&\multirow{2}{*}{\shortstack{PSF\\Frac}}&
    \multirow{2}{*}{Signif}&\multirow{2}{*}{log$P_B$}&\multirow{2}{*}{Anom}&\multirow{2}{*}{Var}&\multirow{2}{*}{\shortstack{ Exp\\(ks)}}&\multirow{2}{*}{\shortstack{$E_{m}$\\(keV)}}&\multirow{2}{*}{\shortstack{Hardness\\Ratio}}\\
    \cmidrule(r){1-2}
    Label&CXOU J\\
    (1)&(2)&(3)&(4)&(5)&(6)&(7)&(8)&(9)&(10)&(11)&(12)&(13)&(14)&(15)&(16)&(17)&(18)
} 
\startdata
1  &185629.72+020437.9&284.12384&2.077201&0.7&10.2&36.1&9.6&459.0&22.4&0.9&3.7&\textless-5&&a&150.3&2.3&0.40\\
2  &185634.65+020942.7&284.144379&2.161875&0.6&8.7&25.7&7.5&215.0&24.1&0.9&3.3&-4.6&g&a&179.9&3.9&0.65\\
3  &185635.12+020729.6&284.146362&2.12491&0.5&8.4&39.5&8.4&191.0&36.4&0.9&4.6&\textless-5&&a&179.9&4.6&0.73\\
4  &185635.12+021528.3&284.146362&2.257876&1.5&11.7&13.3&5.3&101.0&10.0&0.9&2.4&-2.8&&a&29.7&4.6&0.46\\
5  &185638.11+020931.5&284.158813&2.158753&0.6&7.8&18.1&6.4&242.0&15.9&0.9&2.7&-3.3&&c&179.9&3.6&0.59\\
6  &185639.39+020837.3&284.164154&2.143698&0.4&7.4&36.9&7.7&109.0&11.5&0.9&4.7&\textless-5&&b&179.9&1.6&-0.05\\
7  &185643.95+021331.6&284.183167&2.225449&0.7&8.5&17.2&7.1&283.0&12.2&0.9&2.3&-2.4&&a&179.9&2.4&0.49\\
8  &185643.98+020655.3&284.183258&2.115373&0.3&6.2&52.8&7.9&114.0&42.8&0.9&6.7&\textless-5&g&c&179.9&3.5&0.61\\
9  &185644.00+020245.2&284.18335&2.045914&0.4&7.9&49.2&8.6&100.0&17.8&0.9&5.6&\textless-5&g&a&179.9&1.6&-0.04\\
10 &185645.63+021105.6&284.190155&2.1849&0.5&6.7&17.1&5.8&122.0&2.6&0.9&2.8&-3.6&&a&179.9&1.6&-0.12\\
\enddata
\tablecomments{Basic information of sources in G35.4 is presented here:
Col. (1): X-ray catalog sequence number, sorted by right ascension. 
Col. (2): IAU designation. 
Cols. (3) and (4): Right ascension and declination for epoch J2000.0.
Col. (5): Estimated random component of position error, $1\delta$, computed as (standard deviation of PSF inside extraction region)/(number of counts extracted)$^{1/2}$. 
Col. (6): Offaxis angle. 
Cols. (7) and (8): Estimated net counts extracted in the total energy band (0.5-8 keV ) and average of the upper and lower $1\delta$ errors on col. (7). 
Col. (9): Background counts extracted (total band). 
Col. (10): Estimated net counts extracted in the hard energy band (2-8 keV ). 
Col. (11): Fraction of the PSF (at 1.497 keV ) enclosed within the extraction region. Note that a reduced PSF fraction (significantly below 90\%) may indicate that the source is in a crowded region. 
Col. (12): Photometric significance computed as (net counts)/(upper error on net counts). 
Col. (13): Log probability that extracted counts (total band) are solely from background. Some sources have PB values above the 1\% threshold that defines the catalog because local background estimates can rise during the final extraction iteration after sources are removed from the catalog. 
Col. (14): Source anomalies: g = fractional time that the source was on a detector ( FRACEXPO from mkarf) is <0.9; e = source on field edge; p = source piled up; s = source on readout streak.
Col. (15 ): Variability characterization based on K-S statistic (total band): a = no evidence for variability ($0.05<P_{K-S}$); b = possibly variable ($0.005<P_{K-S}<0.05$); c = definitely variable ($P_{K-S}<0.005$). No value is reported for sources with fewer than 4 counts or for sources in chip gaps or on field edges. 
Col. (16 ): Effective exposure time: approximate time the source would have to be observed on axis to obtain the reported number of counts. 
Col. (17): Background-corrected median photon energy (total band).
Col. (18): Hardness Ratio defined as $\frac{H-S}{H+S}$, where H and S correspond to source counts of hard band and soft band.
Table 4 is available in its entirety in its machine readable format. 
A portion is shown here for guidance regarding its form and content.}
\end{deluxetable*}
\clearpage

\subsection{Source Variability}\label{sec:variability}

\begin{figure*}
\gridline{\fig{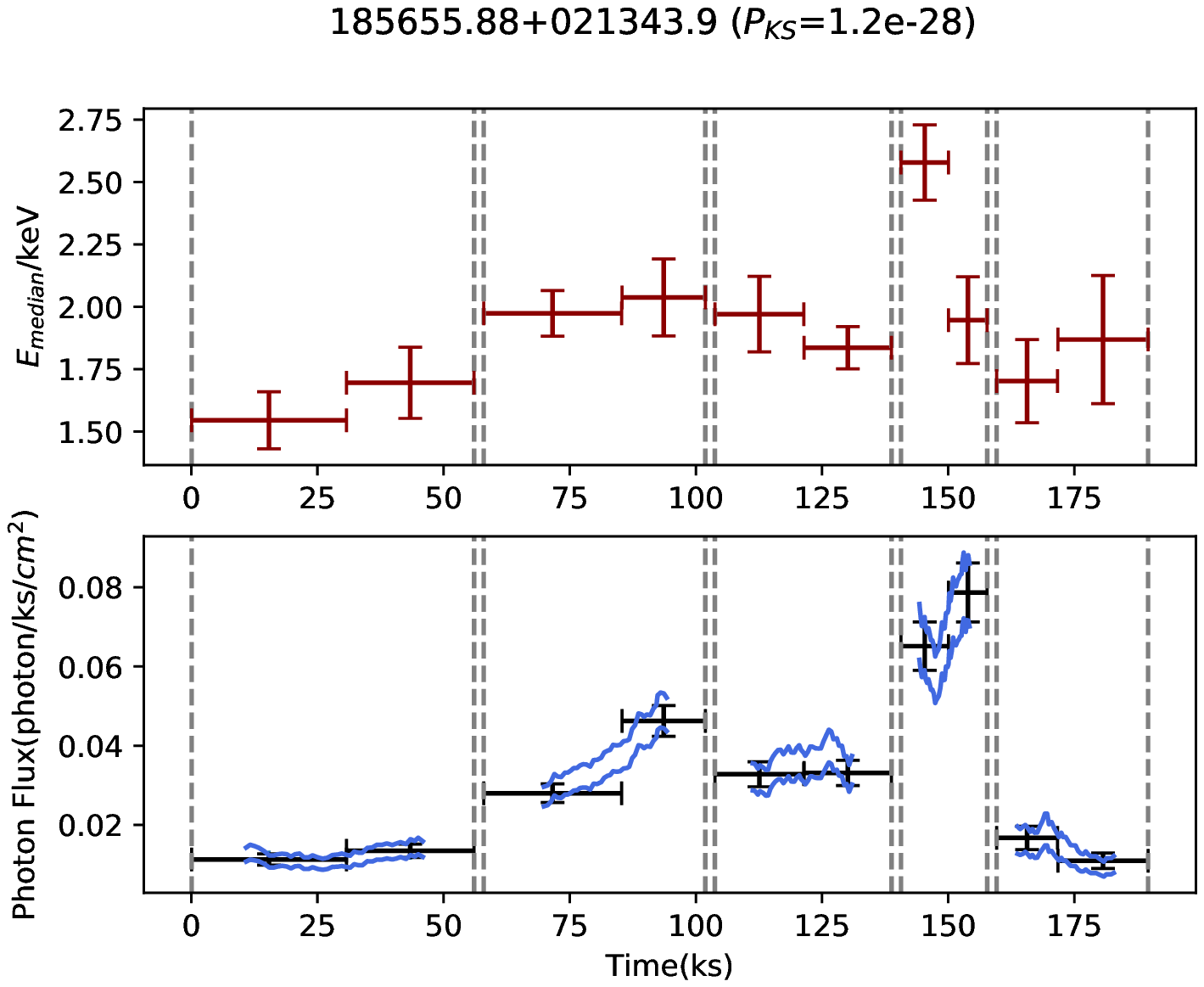}{0.5\textwidth}{(a)}
          \fig{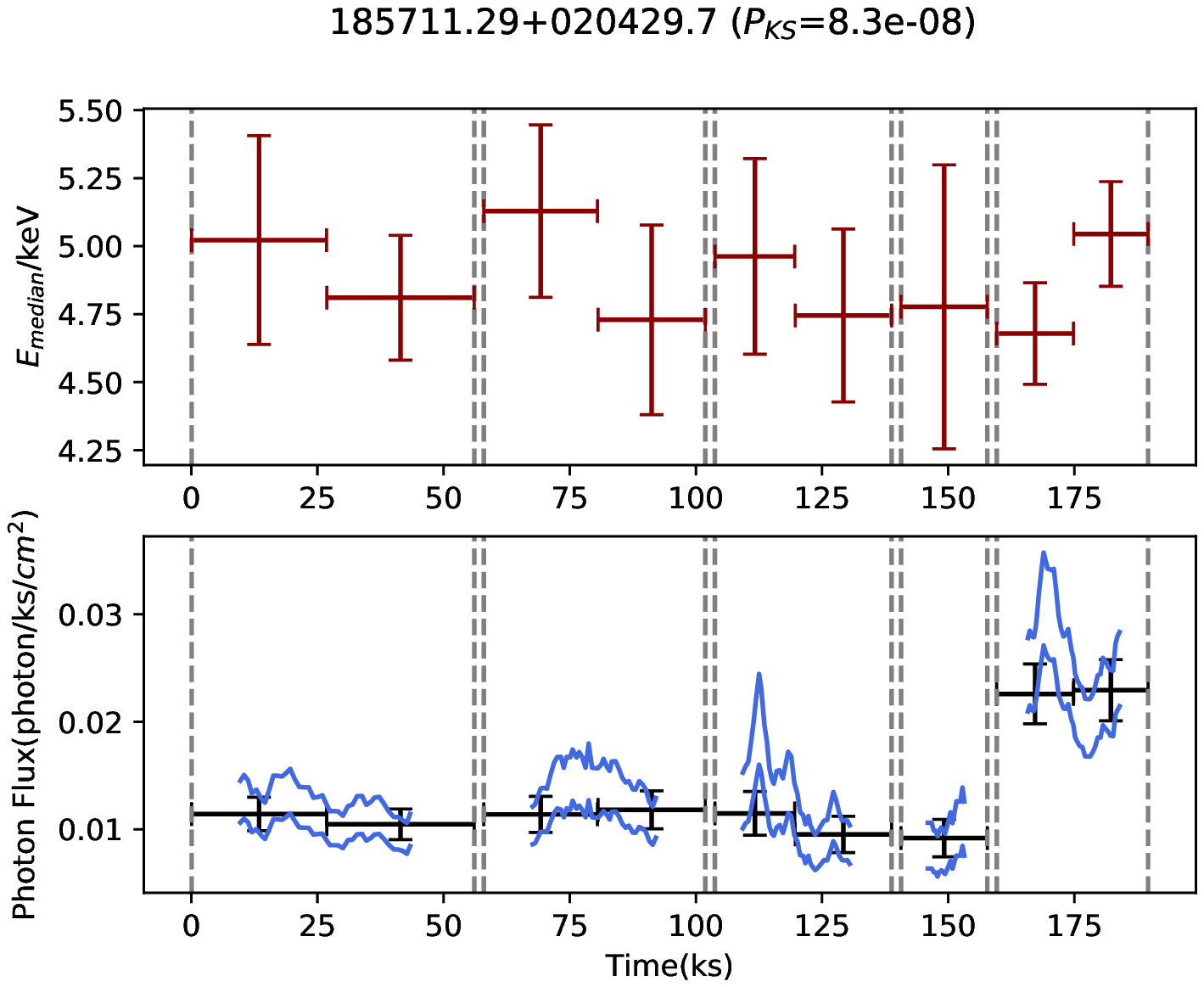}{0.5\textwidth}{(b)}
          }
\gridline{
          \fig{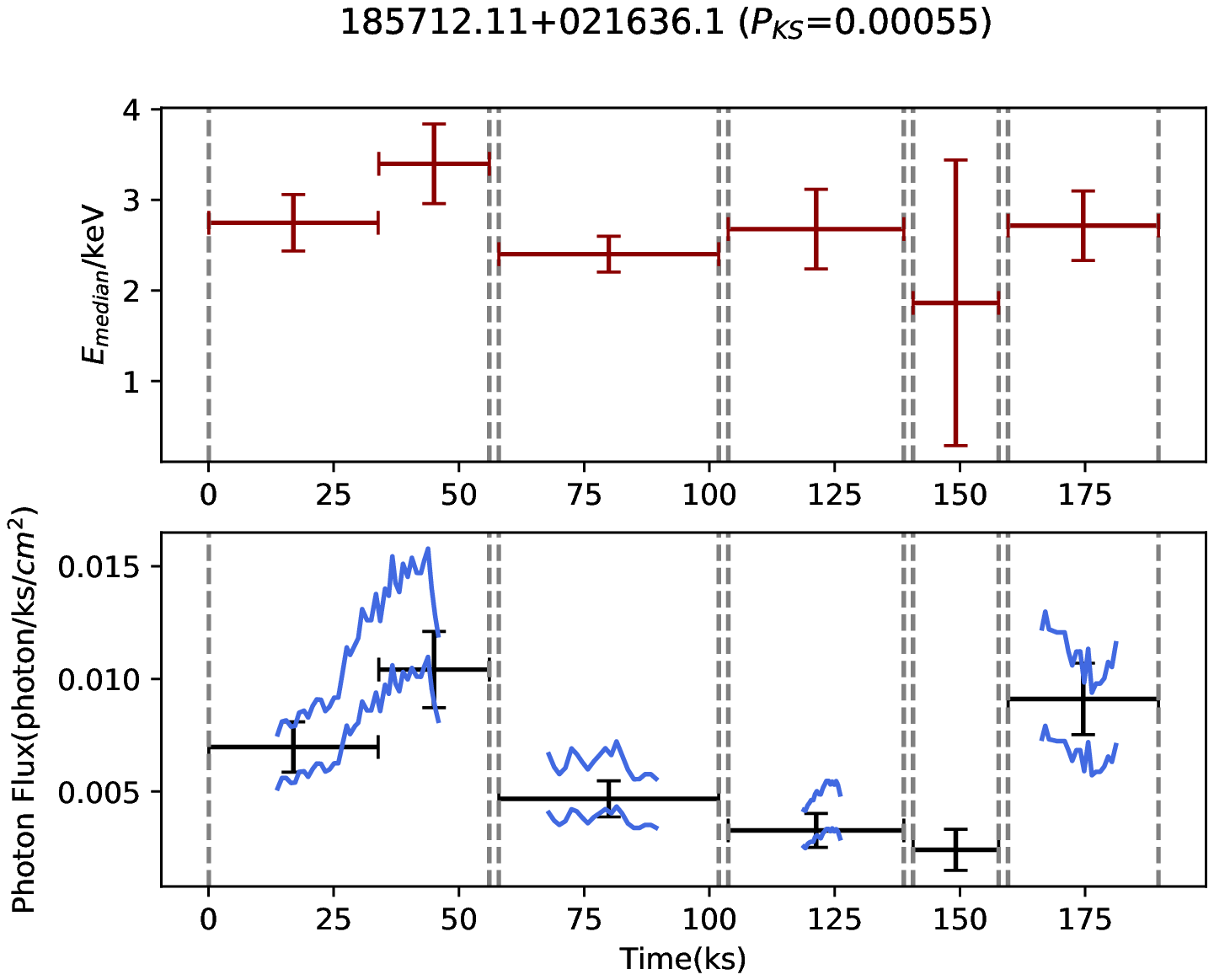}{0.5\textwidth}{(c)}
          }
\caption{
Flux and median energy time series for multiple observations of
(a) 185655.88+021343.9, (b) 185711.29+020429.7, and
(c) 185712.11+021636.1.  The median energies of the events in the light
curve bins are shown in red, while the binned light curve is shown in
black. Poisson errors within a single time bin are
represented by black and red error bars along the y-axis.  The pair of
blue curves present the 68\% confidence intervals for the
continuous light curve, which is estimated from X-ray events using an
adaptively-sized sliding window. Note occasionally binned data points (e.g., at $\sim$50 ks and 120 ks in (c)) are close to the boundary of confidence intervals. In the case of a brief flare during a time bin, the continuous curves give a more accurate representation of the variability with respect to the flux sampled with large time bins.
\label{fig:lightcurve}}
\end{figure*}

It is one of the most notable characteristics of PMS stars that they flare in the X-ray band, and a few extremely powerful X-ray flares have been
reported in several PMS
stars\citep[e.g.][]{Imanishi,Grosso,Favata,getman06,jfw07,broos07}.
AE scripts is used to perform a Kolmogorov-Smirnov (K-S) test for each
X-ray source to assess light curve variability during these
observations. The K-S test compares the distribution of photon arrival time stamps with that expected for a constant rate source. We find that 10 sources of G34.4 showed significant
variability ($P_{KS}<0.005$), but all of them have no more than 100 net counts.  We also identify 20 sources of G35.4 that showed significant
variability, and 3 of these emitted more than 100 net counts.  Light curves of these three sources are shown in Figure \ref{fig:lightcurve}.
Due to the short duration of each individual observation, no entire flare, 
which is supposed to include both the rising and the decaying part, was recorded. 
And it is informed by the light curves that none of these flares is as intense as the super-flares seen in the Orion population and the Cep B region \citep{getman06}.  One source, CXOU J185655.88+021343.9, shows a large amplitude variation (a factor of 8) flare across the multiple observations, and the median energy of the photons become higher during the peak of the flare. Its X-ray luminosity in the 0.5--8 keV band is log~$L_X=30.43$ erg s$^{-1}$. This is typical behavior of the flares in known PMS stars in Orion \citep{Grosso,Favata}. It is also interesting to note that in its light curve (Figure~\ref{fig:lightcurve}a), the peak of the median energy (at $\sim$140 ks) does not correspond to the peak of the flux. We find this also exists in the flaring characteristics of other young stellar objects (see Table 1 in \citet{2008ApJ...688..418G} for example). It is plausible the reconnected flaring loop structure is still expanding whilst cooling is already significant, resulting in a lag between the peaks of spectral hardness and flux \citep{Favata,2010ARA&A..48..241B}. An alternative scenario is that after the initial heating even when the temperature start decreasing, the emission measure still increases due to an increasing plasma density in the flaring loop structure contributed by the evaporating plasma from the chromosphere filling the loop through its feet \citep{Reale07}.

\subsection{Spectral Fitting} \label{sec:fitting}

The spectral fitting process is carried out only for valid sources
with more than 10 net counts. These spectra are fitted with three
different types of models. Two of these are thermal plasma models \citep{apec} typically adopted for X-ray emission of PMS stars. One with a single temperature and the other with two
temperature components, specifically the \emph{vapec} and \emph{vapec+vapec} models included in the  \emph{XSPEC} package \citep[version 12.10.0;][]{Arnaud96}\footnote{More information can be
found at \url{https://heasarc.gsfc.nasa.gov/xanadu/xspec/} and \url{https://atomdb.org}}. The \emph{powerlaw} model has a power law spectrum
of photon energies, whose energy distribution follows 
\begin{equation}
A(E)=KE^{-\alpha},
\end{equation}
where the parameter $\alpha$ refers to the photon index of this power-law distribution, and $K$ refers to a constant of $\rm photons/keV/cm^2/s$ at 1 keV. 

\begin{deluxetable*}{cccccccccccc}
\tablenum{4}
\tablecaption{Spectral fitting results of X-ray sources with more than 100 net counts in the full band\label{tab:manyfit}}
\tablewidth{0pt}
\tablehead{
    \\
    \multirow{3}{*}{Target}&\multirow{3}{*}{Label}&\multicolumn{4}{c}{Spectral Fit}&\multicolumn{5}{c}{X-ray Luminosities}&\multirow{3}{*}{Notes}\\
    \cmidrule(lr){3-6}
    \cmidrule(lr){7-11}
    &&\multirow{2}{*}{\shortstack{$logN_H$\\($cm^{-2}$)}}&\multirow{2}{*}{$\chi^2$}&\multicolumn{2}{c}{\multirow{2}{*}{\shortstack{Model\\Parameters}}}&\multirow{2}{*}{\shortstack{$logL_s$\\($ergs\ s^{-2}$)}}
    &\multirow{2}{*}{\shortstack{$logL_h$\\($ergs\ s^{-2}$)}}
    &\multirow{2}{*}{\shortstack{$logL_{h,c}$\\($ergs\ s^{-2}$)}}
    &\multirow{2}{*}{\shortstack{$logL_t$\\($ergs\ s^{-2}$)}}
    &\multirow{2}{*}{\shortstack{$logL_{t,c}$\\($ergs\ s^{-2}$)}}\\
    \\(1)&(2)&(3)&(4)&\multicolumn{2}{c}{(5)}&(6)&(7)&(8)&(9)&(10)&(11)
}
\startdata
\multirow{2}{*}{IRDC G34.4+0.23 }&\multicolumn{2}{c}{tbabs\_pow}&&$\Gamma$&$logN_{\Gamma}$\\
\cmidrule(lr){2-6}
&107&$22.5_{+0.2}^{-0.2}$&0.72&$0.4^{+0.4}$&$-5.3^{+0.3}_{-0.5}$&30.3&32.3&32.3&32.3&32.4\\   
\hline
\multirow{14}{*}{IRDC G35.4-0.33}&\multicolumn{2}{c}{tbabs\_vapec}&&$kT$(keV)&$logEM$\\
\cmidrule(lr){2-6}
&11 &$21.0^{+0.7}$&1.82&$0.5^{+0.1}_{-0.1}$&$54.1^{+0.2}_{-0.3}$&30.6&30.2&30.2&30.8&30.9\\
&44 &$21.7^{+0.1}_{-0.1}$&1.05&$4.1^{+0.6}_{-0.6}$&54.7&31.4&31.7&31.8&31.9&32.1\\
&$96^*$&$22.8^{+0.1}_{-0.1}$&0.74&$5.4^{+3.2}_{-3.1}$&$55.0^{+0.1}_{-0.2}$&29.6&31.7&31.9&31.8&32.1\\
&129&$22.3^{+0.1}_{-0.1}$&0.38&$4.5^{+3.0}_{-3.0}$&$54.2^{+0.2}_{-0.2}$&30.3&31.2&31.3&31.2&31.6&Class $\uppercase\expandafter{\romannumeral2}$\\
&160&$23.0^{+0.2}_{-0.1}$&0.45&$9.5^{+18.4}$&$54.5^{+0.2}_{-0.5}$&30.0&31.4&31.6&31.4&31.7\\
\cmidrule(){2-12}
&\multicolumn{2}{c}{tbabs\_pow}&&$\Gamma$&$logN_{\Gamma}$\\
\cmidrule(lr){2-6}
&24 &$22.7^{+0.1}_{-0.2}$&0.23&$10.0^{+0.3}_{-0.3}$&$-5.4^{+0.1}_{-0.3}$&29.8&32.1&32.2&32.1&32.2\\
&50 &$21.0^{+0.7}$&0.73&$3.1^{+0.5}_{-0.4}$&$-5.1^{+0.1}_{-0.2}$&31.1&30.6&30.6&31.2&31.3\\
&123&$23.4_{-0.1}$&0.83&$1.4^{+0.5}_{-0.5}$&$-4.3^{+0.3}_{-0.6}$&27.7&32.1&32.4&32.1&32.6\\
&157&$22.6^{+0.2}_{-0.2}$&0.85&$1.0^{+0.5}_{-0.6}$&$-5.5^{+0.4}_{-0.8}$&29.8&31.4&31.5&31.4&31.6\\
\cmidrule(){2-12}
&\multicolumn{2}{c}{tbabs\_2vapec}&&$kT$(keV)&$logEM$\\
\cmidrule(lr){2-6}
&\multirow{2}{*}{176}&\multirow{2}{*}{$21.5^{+0.3}_{-1.1}$}&\multirow{2}{*}{0.60}&$0.4^{+0.1}_{-0.1}$&$54.2^{+0.2}_{-0.6}$&\multirow{2}{*}{31.0}&\multirow{2}{*}{30.3}&\multirow{2}{*}{30.3}&\multirow{2}{*}{31.0}&\multirow{2}{*}{31.1}\\
&&&&$9.5^{+67.8}$&$53.3^{+0.3}$\\
\enddata
\tablecomments{Spectral fitting result of X-ray sources with more than 100 net counts is presented here:
Cols. (1) and (2): the target and the label of the source in the source list of the target. 
Cols. (3) and (4): Estimated column density and reduced $\chi^2$ for the spectra fit.
Cols. (5): Model parameters of sources, which is plasma energy and plasma emission measure for sources of \emph{tbabs\_vapec} and \emph{tbabs\_2vapec} models, while power-law photon index parameter ($\Gamma$) and powerlaw normalization ($N_{\Gamma}$) for \emph{tbabs\_pow} model.
Cols. (6)-(10):Observed and corrected for absorption X-ray luminosities, obtained from spectral analysis.
\\
$*-$ An alternative model consisting of a powerlaw continuum and a gaussian emission line also provides satisfactory fit, with $\chi^2=0.73$. The energy of line center is $E_0=6.9\pm 0.1$ keV and the equivalent width $EW=1.1^{+0.7}_{-0.5}$ keV.}
\end{deluxetable*}

In general, unconstrained fits are preferred (including power law fits) and at first the single-temperature thermal plasma model is the default model used for spectral fitting. If a one-temperature thermal plasma model did not fit the data well, a two-temperature thermal plasma model or a power law model was invoked.
To evaluate the best-fit model, the maximum likelihood tests were run.  A $\chi^2$-statistic was adopted for sources with more than 100 counts, while a $c$-statistic (Cash 1979) was used for those fainter sources with between 10 and 100 counts.  In the situation where the likelihood of the single temperature model and two-temperature thermal plasma model were similar and the lower temperature of the two temperature model is unphysically low ($\ll 0.1$~keV), the single \emph{vapec} model was retained.

Spectral analysis results for sources with net counts more than 100 in
the full band are presented in Table \ref{tab:manyfit}.  Fit results
for other sources with fewer net counts are shown in Table
\ref{tab:g344fit} and Table \ref{tab:g354fit}.  The table notes give
detailed descriptions of the columns. For sources of G35.4 that are expected to be Class \uppercase\expandafter{\romannumeral2} sources, the
median of their absorbing column densities is $\log N_{\rm H}/{\rm cm}^{-2} \simeq 22.3$, and the median of their temperatures is about 2.9 keV. Such
column densities and temperatures are similar to those found in
previous studies of X-ray emission from deeply embedded PMS stars
\citep{getman06,jfw07,povich16,gunther12}.

\section{Identification of Chandra source counterparts}

\begin{figure*}
\gridline{\fig{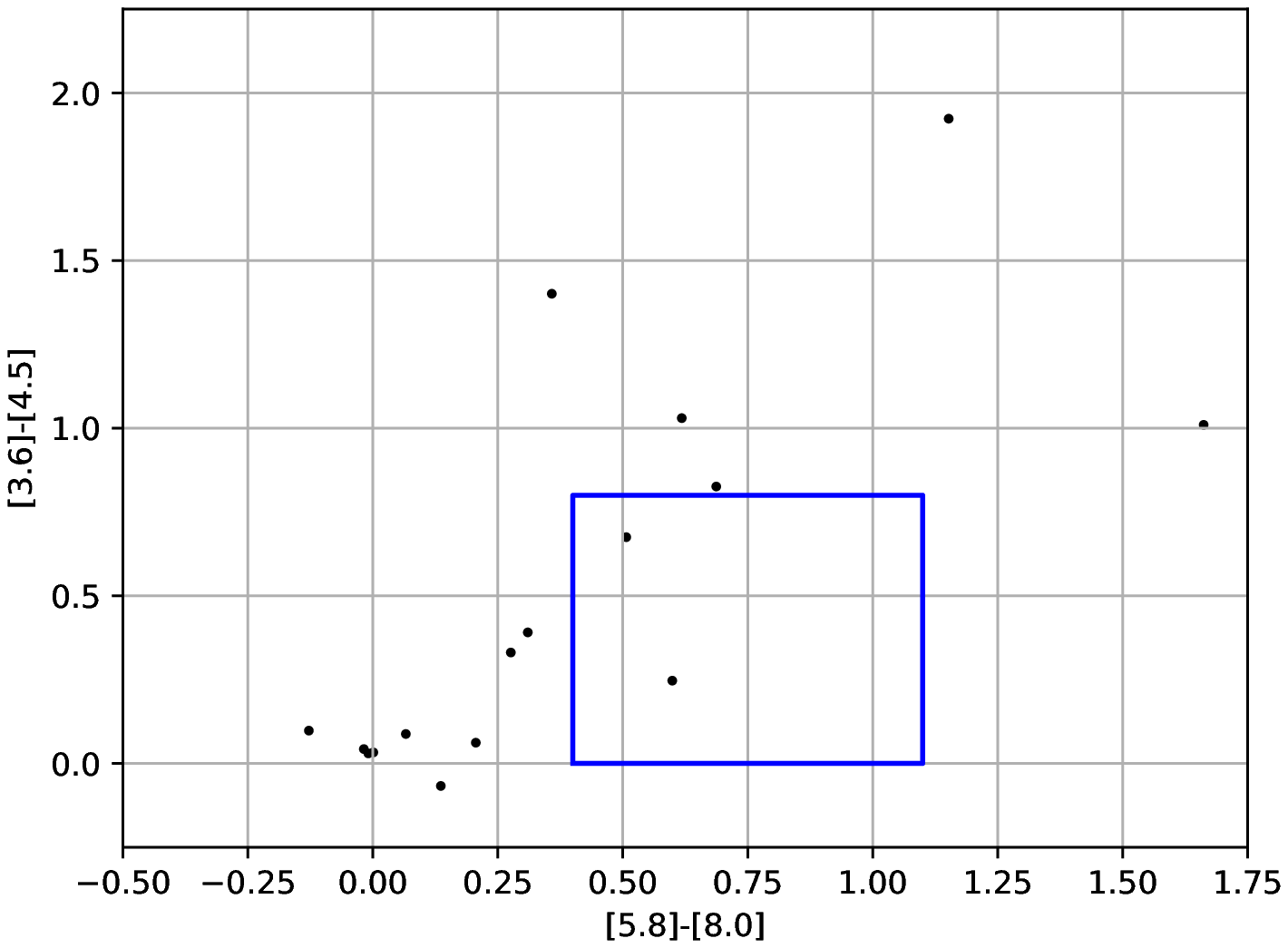}{0.45\textwidth}{(a)}
          \fig{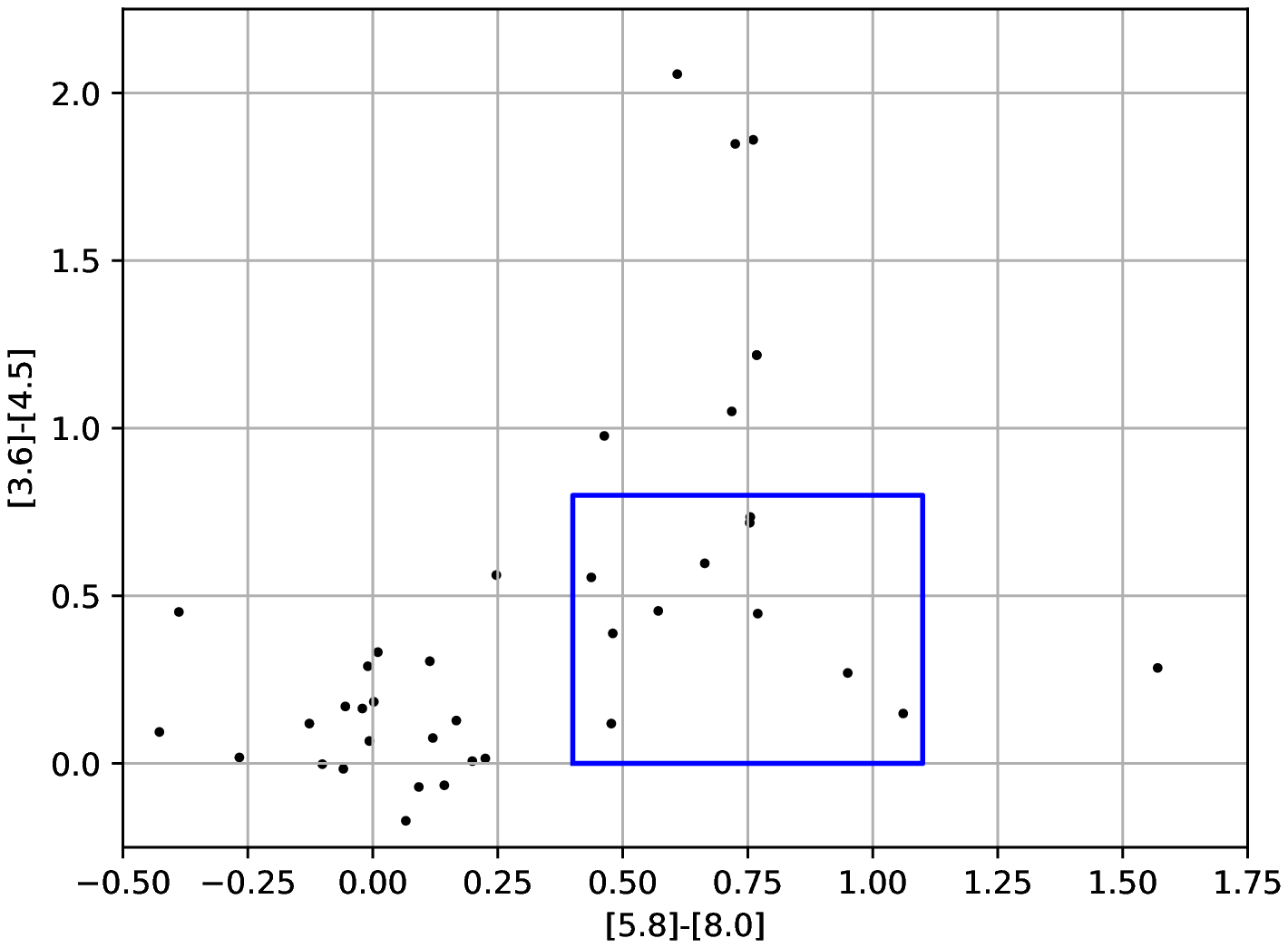}{0.45\textwidth}{(b)}
          }
\gridline{
          \fig{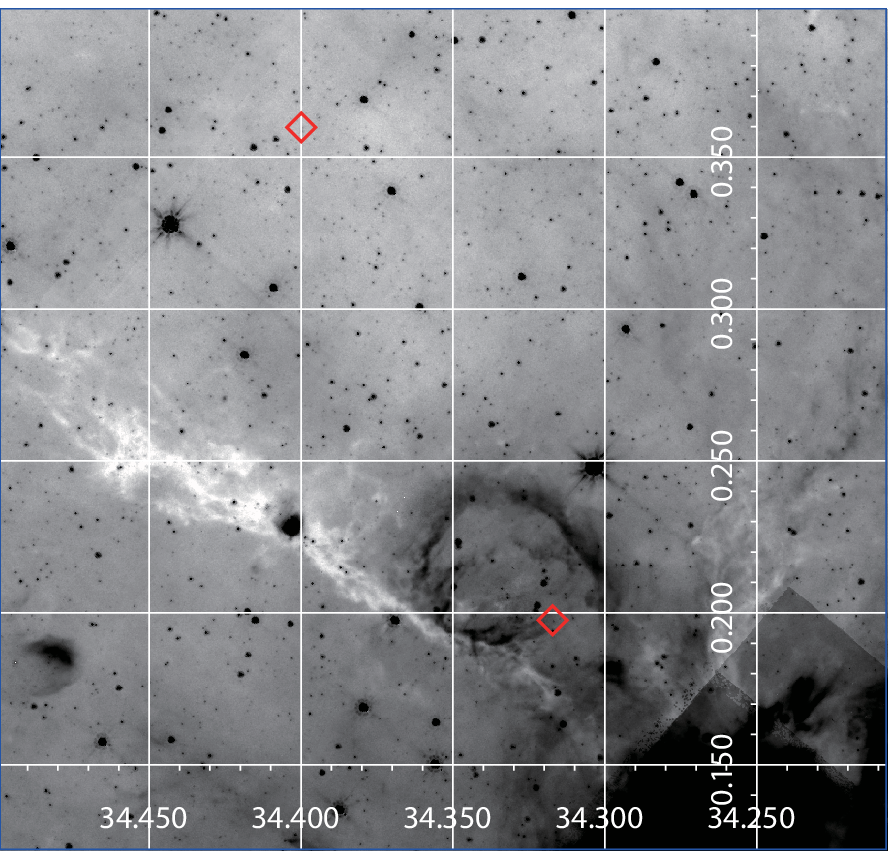}{0.45\textwidth}{(c)\spitzer{} image of G34.4}
          \fig{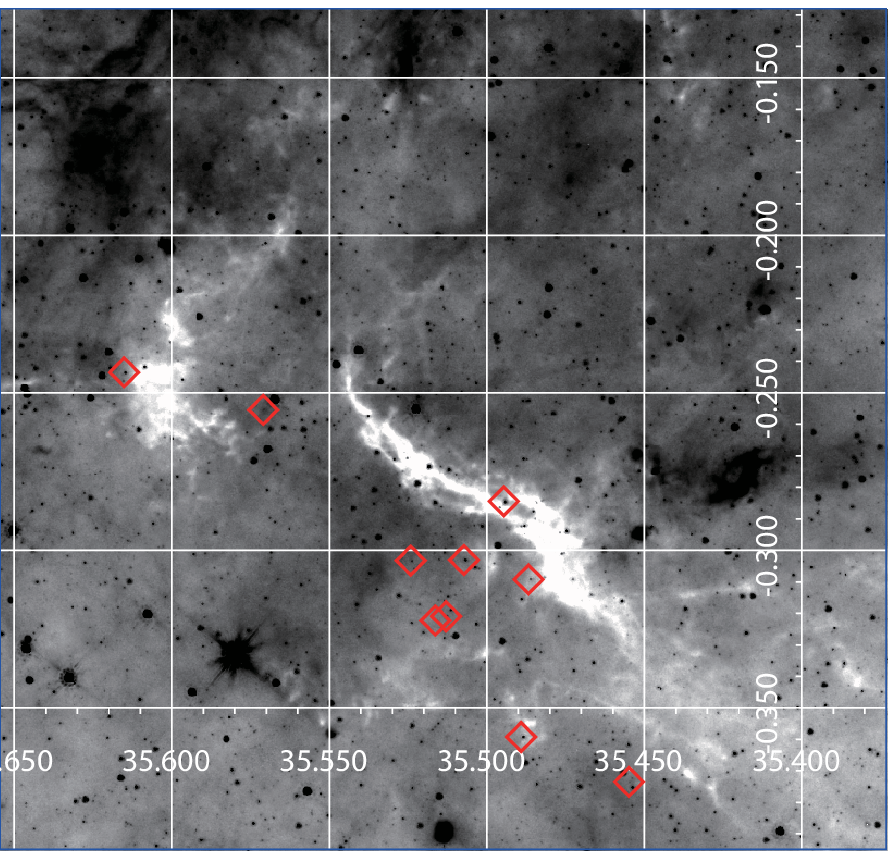}{0.45\textwidth}{(d)\spitzer{} image of G35.4}
          }
\caption{
Panels (a) and (b) are \spitzer{} [3.6]-[4.5] versus [5.8]-[8.0]
color-color diagrams for the IRDCs G34.4 and G35.4, respectively, generated
following \citet{lori}.  The blue rectangular area is typically occupied by Class
\uppercase\expandafter{\romannumeral2} sources. 
Panels (c) and (d) are inverted grey scale \spitzer{} $8\:{\rm \mu m}$ images showing the locations of these Class \uppercase\expandafter{\romannumeral2} sources identified in the
color-color diagrams shown as red diamonds.  
\label{fig:spitzer}}
\end{figure*}

The X-ray sources identified in Section 2 were then assessed for
association with near-IR sources via a positional coincidence criterion. We cross-matched the X-ray sources to those in
the infrared archives: 2MASS \citep[Two Micron All Sky Survey,][]{2mass};
WISE (Wide-field Infrared Survey Explorer) All-Sky Data Release
\citep{wise}; and GLIMPSE \citep[Galactic Legacy Infrared Midplane
  Survey Extraordinaire using the \spitzer{} Space Telescope,
][]{spitzer,glimpse}.
The \chandra{} sources are matched to the 2MASS and GLIMPSE catalog of
point sources with a 0.5\arcsec\ matching radius, and to WISE with a
2\arcsec.  The VizieR (\url{http://vizier.u-strasbg.fr/}) catalog
services are applied for complimentary information.  The results of
this cross-matching are reported in Table \ref{tab:g344sc} and Table
\ref{tab:g354sc} for the two IRDCs.

Photometric information of matched  \spitzer{} GLIMPSE counterparts was utilized to classify the type of these
sources based on their mid-IR colors.  Following the method in \citet{lori}, color-color diagrams (Figure
\ref{fig:spitzer}) were made by employing these data products.  Two rectangular regions in these figures approximate the domain of Class
\uppercase\expandafter{\romannumeral2} sources, and their locations
are shown in the \spitzer{} image. Especially in IRDC G35.4, it is apparent that
the identified Class II sources are often close to IR extinction
features where dense gas is, although they tend to be absent in the darkest
regions where the obscuration column density is too high to reveal any embedded X-ray sources.

\begin{figure*}
\gridline{\fig{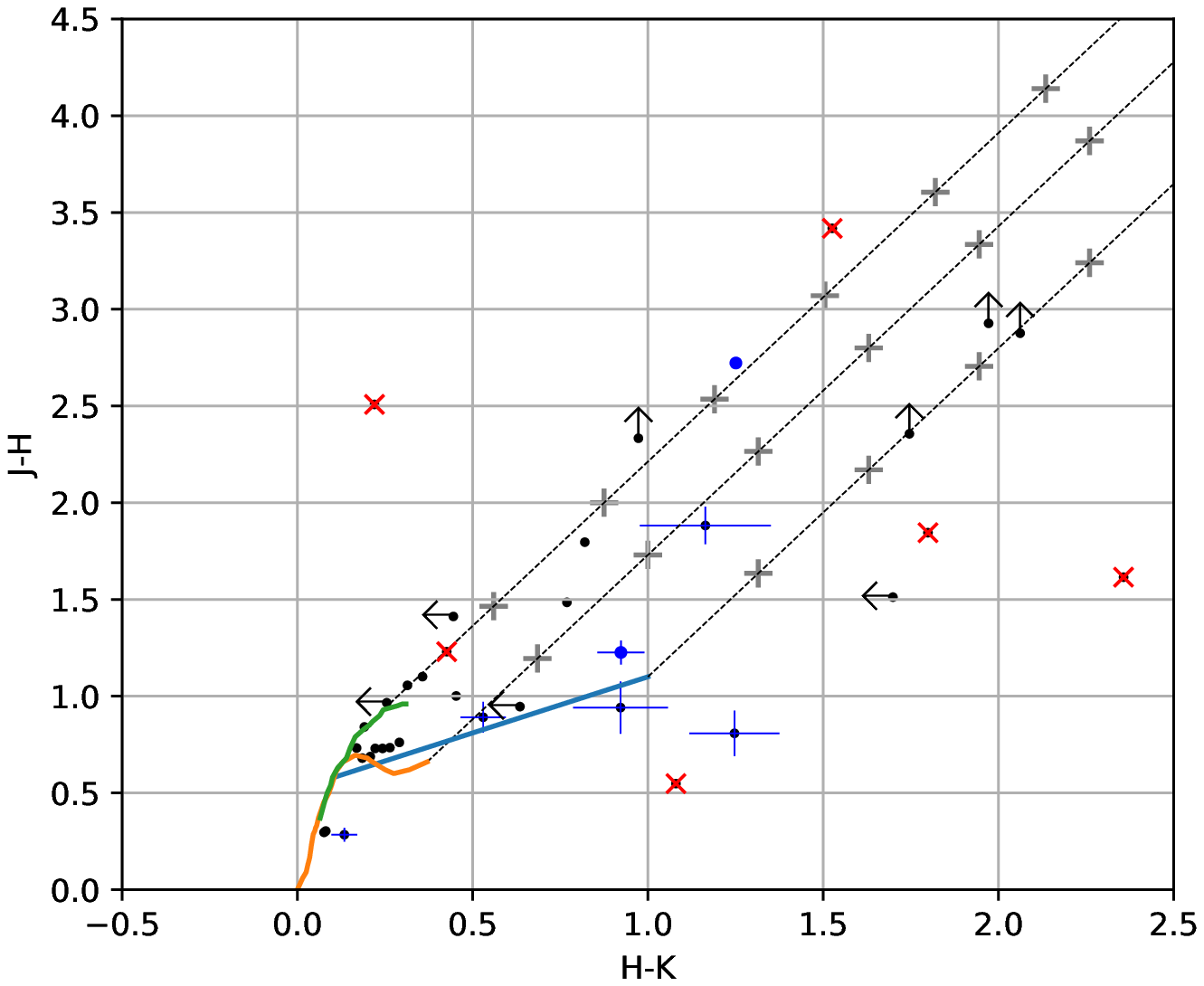}{0.45\textwidth}{(a)}
          \fig{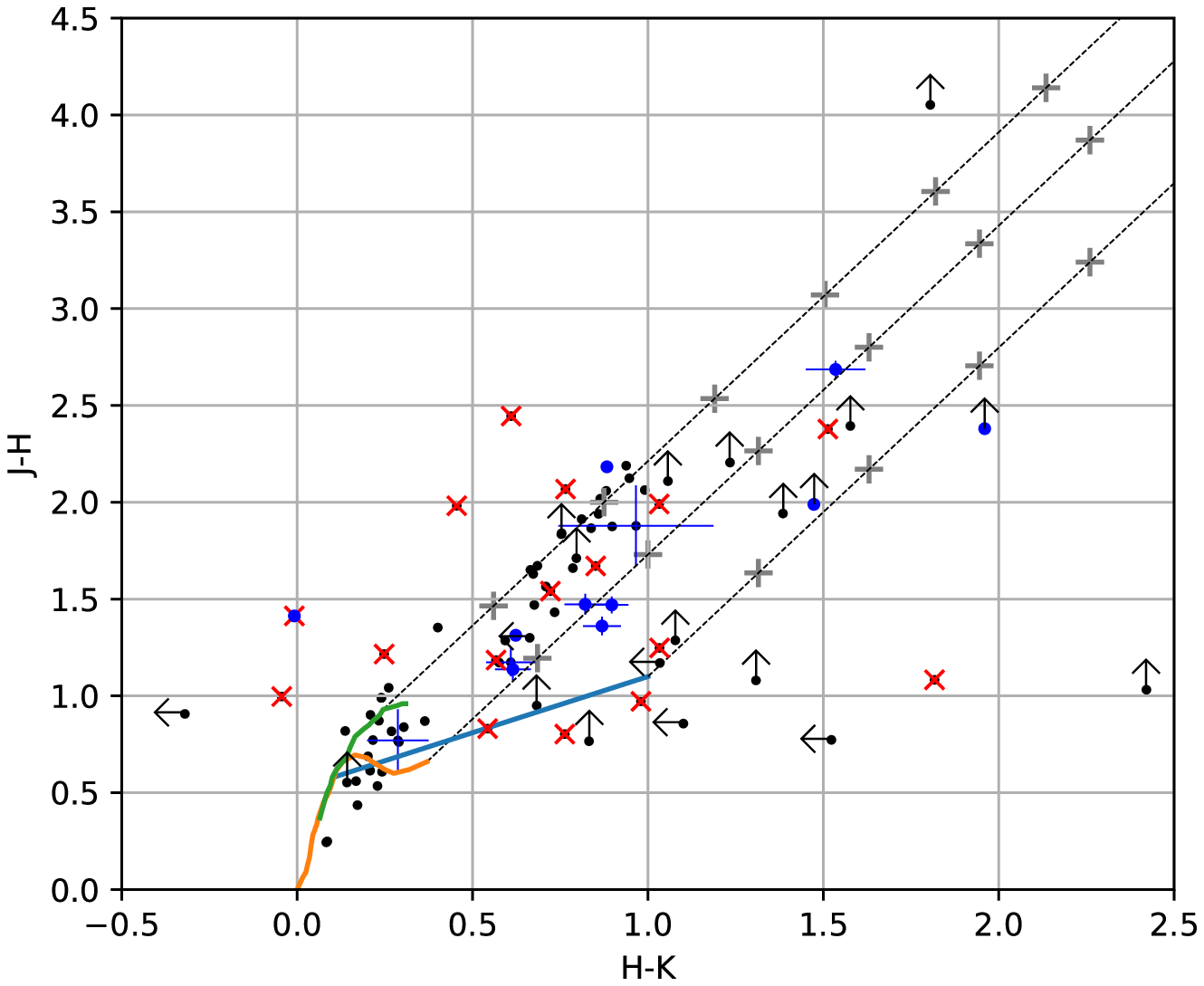}{0.45\textwidth}{(b)}
          }
\gridline{\fig{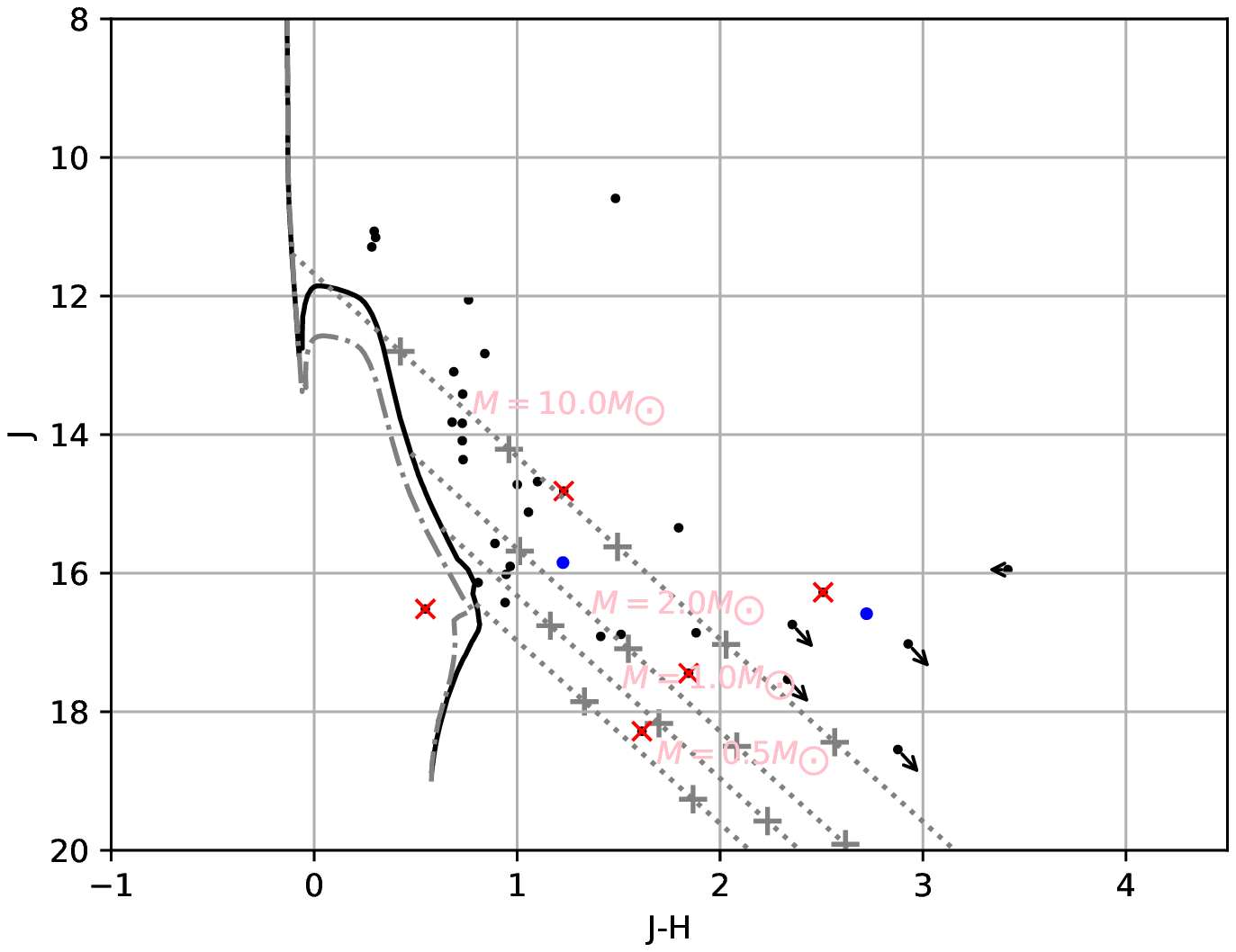}{0.45\textwidth}{(c)}
          \fig{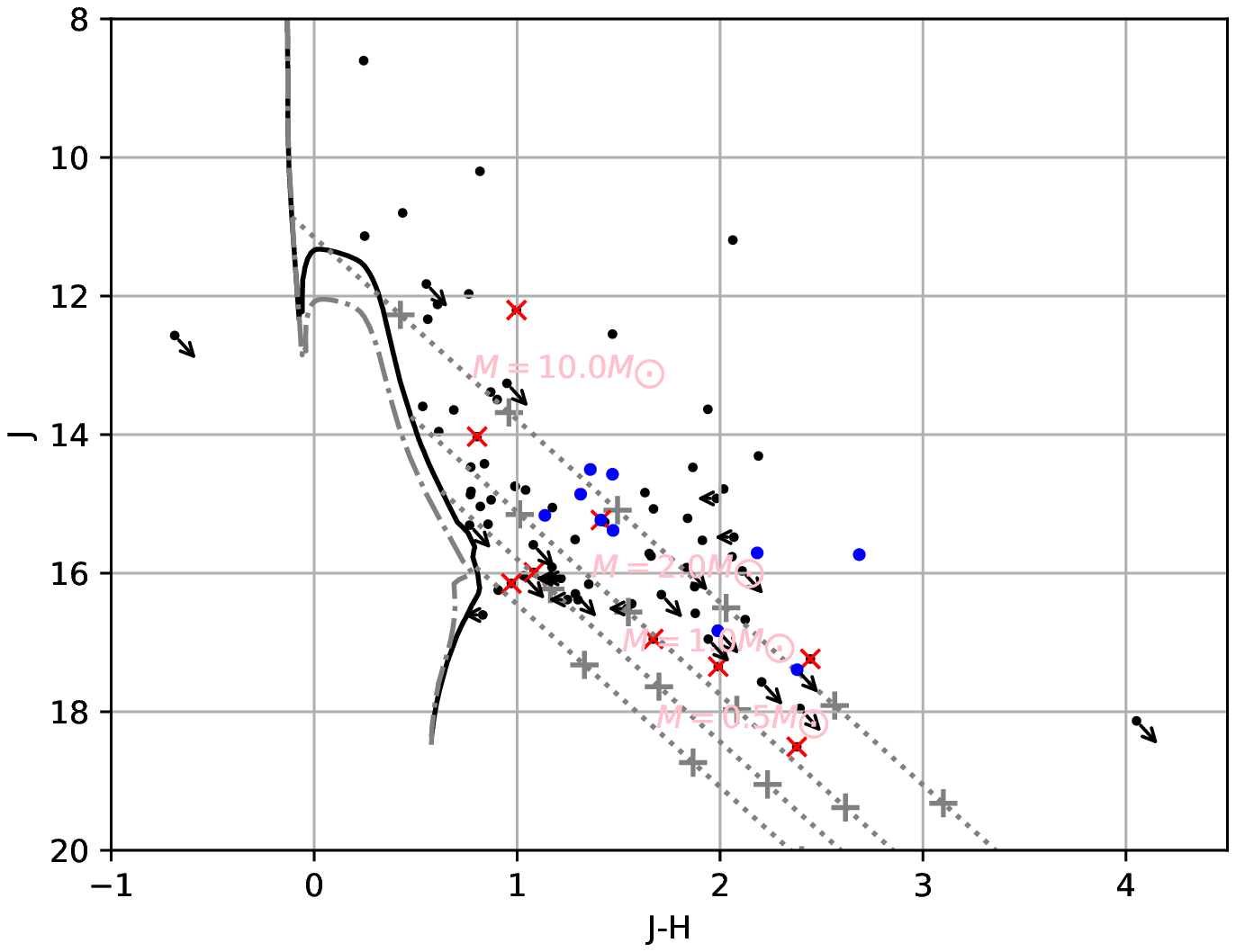}{0.45\textwidth}{(d)}
          }
\caption{ 
Panels (a) and (b) are $J-H$ versus $H-K$ color-color diagram for G34.4
and G35.4, respectively.  The green line and yellow line in each
diagram represents the locus for Main Sequence stars and giants from
\citet{bb1988}, respectively. 
The blue line in each diagram is the locus of CTTS from \citet{meyer1997}.
Sources with photometric measurement as upper limit in one band were marked as arrows, and each of these arrows shows the associated limit on the color.
Sources with upper limit in more than one band is marked as a cross, which means their positions on this diagram are uncertain. 
Panels (c) and (d) are J versus J-H color-magnitude diagrams for G34.4
and G35.4, respectively. The solid and dashed lines are the isochrone tracks of
1 Myr and 2 Myr old PMS, respectively. These evolutionary tracks are based on the PARSEC models (release v1.2S, \citet{cmd,cmd1}) with metallicity of Z=0.0152, available at \url{http://stev.oapd.inaf.it/cgi-bin/cmd/}.  The sources with IR
photometry $J>17$ are less reliable in photometry, and most of them were close to the detection limit. 
Class \uppercase\expandafter{\romannumeral2} identified from Figure
\ref{fig:spitzer} are shown as blue dots in these color-color
diagrams and color-magnitude diagrams. 
The grey dashed lines in four panels represent the standard reddening vectors with grey crosses separated by every $A_V = 5\:$mag of extinction. 
\label{fig:twomass}}
\end{figure*}

The NIR data have long been proven useful to study the nature of stars in young stellar clusters \citep{H92,Lada96,meyer1997}. It has been previously demonstrated that photometric measurements of the matched counterparts from 2MASS could be used to estimate the ages and masses of the YSOs based on theoretical stellar isochrones \citep[e.g.,][]{jfw08}.   The J-H versus H-K color-color diagrams and J
versus J-H diagrams are shown in Figure \ref{fig:twomass}.  Note that sources with photometric measurement as upper limit in 2MASS observation (flagged as 'U' in the photometric quality) were also shown but their locations could be shifted from their original positions in these plots.  In Figures \ref{fig:twomass}a and b, sources located between the two dash lines of reddening through normal interstellar extinction of an O dwarf and an M giant were considered as diskless young stars, whose spectral energy distribution is consistent with reddened photosphere (classified as Class \uppercase\expandafter{\romannumeral3} or Weak-lined T Tauri Stars, WTTS).  Sources located at the right side of this region are K-band excess sources (classified as Class \uppercase\expandafter{\romannumeral2} or Classical T Tauri Stars, CTTS), which were identified in colors as ${(J-H)<1.7(H-K)+2\delta(H-K)}$ \citep{jfw07}. Such K-band excess is thought to be evidence of stars with hot inner disks, where dust might be relatively hot ($T\sim1200K$) and thus have significant dust emission in the near-IR.  The CTTS loci from \citet{meyer1997} is also shown. The distribution of IR sources in G34.4 is too sparse, but the location of sources with photometric measurements in G35.4 implies they are consistent with being Class \uppercase\expandafter{\romannumeral2} and \uppercase\expandafter{\romannumeral3} sources subjected to a range of interstellar reddening.

The $J$ - $H$ versus $J$ color-magnitude diagrams (Figure \ref{fig:twomass}c and d) give both the rough mass distribution and the absorption distribution of the ACIS sources with 2MASS counterparts. The isochrones are 1~Myr and 2~Myr evolutionary tracks for PMS stars based on the PARSEC models (release v1.2S, \citet{cmd,cmd1}) with a metallicity of Z=0.0152, shifted with the distance modulus of each target applied (corresponding to 3.7 kpc for G34.4 and 2.9 kpc for G35.4) and marked with reddening vectors for estimation of the dereddened masses and the associated absorption. With a typical reddening $A_V\sim 5-10$ (inferred from Figure\ref{fig:twomass}a and b, also consistent with X-ray absorption column from spectral fitting), the scattered data points fall close to the 1--2 Myr isochrones. Hence we adopted 1--2 Myr for the age of these stars, as a young population is expected in and around the IRDCs. We caution that the estimated mass and the amount of absorption for a given star should not be taken as precise measurement, since they also rely on the distance and age of the cluster besides the uncertainties in the star's photometry.

The faint X-ray sources without NIR counterparts are likely to be background stars or extragalactic sources. If this is the case, they would show higher absorption and hard spectra in general. We have examined both median energy and hardness ratio (listed in Tables~2 and 3) of the X-ray sources, and through comparison we find that the ones without stellar counterparts indeed show higher hardness ratio and median energy than those have.
Also some sources with much larger numbers of counts than the other sources in the ACIS observation could be foreground contaminants. For such foreground sources, the intervening
Galactic hydrogen column density $N_H$ should be relatively low.  In
fact, we find two sources in G35.4 (sequence numbers 11 and 50) with
much more net counts compared to other sources and having quite low
$N_H$ in the spectral fitting result.  One of these sources (sequence
number 11) was confirmed to be 0.13 arcsec away from HD 175746, a bright foreground star.

\section{X-ray Luminosity Function}

It was proposed by \citet{feig05} that the X-ray Luminosity Function (XLF) of a young stellar cluster can be viewed as the convolution of the stellar initial mass function (IMF) and the X-ray luminosity-mass correlation, which has been measured in the well-known COUP studies \citep{preibisch}.  We then examine the XLFs based on spectral fitting, as a way to quantify the distribution of the IMF and estimate the population of young stellar clusters. Such population analysis has been carried out in Cep OB3b
\citep{getman06}, NGC6357 \citep{jfw07} and NGC2244 \citep{jfw08}.  Using the best-studied Orion Nebula Cluster (ONC) XLF as a calibrator, the XLF of G35.4 could be used to probe the IMF and the population of this IRDC.  For G34.4 we evaluated that there are too few luminosity bins above the completeness limit to constrain the slope of XLF.  In the following XLF analysis, we used the absorption-corrected hard band XLF rather than the total-band XLFs, as the unconstrained soft band component of heavily absorbed X-ray sources likely introduces large uncertainties in both observed and absorption-corrected total-band X-ray luminosities.  The counts of sources in different X-ray luminosity bins were used to construct the absorption-corrected hard-band XLF for all sources.  The absorption-corrected hard-band fluxes of each sources here was obtained from XSPEC spectral fitting scripts. The absorption-corrected hard band XLF for COUP is calculated from the COUP catalog \citep{getman05a}, and used for comparison with the XLF of G35.4 in the same band.

\begin{figure}
    \fig{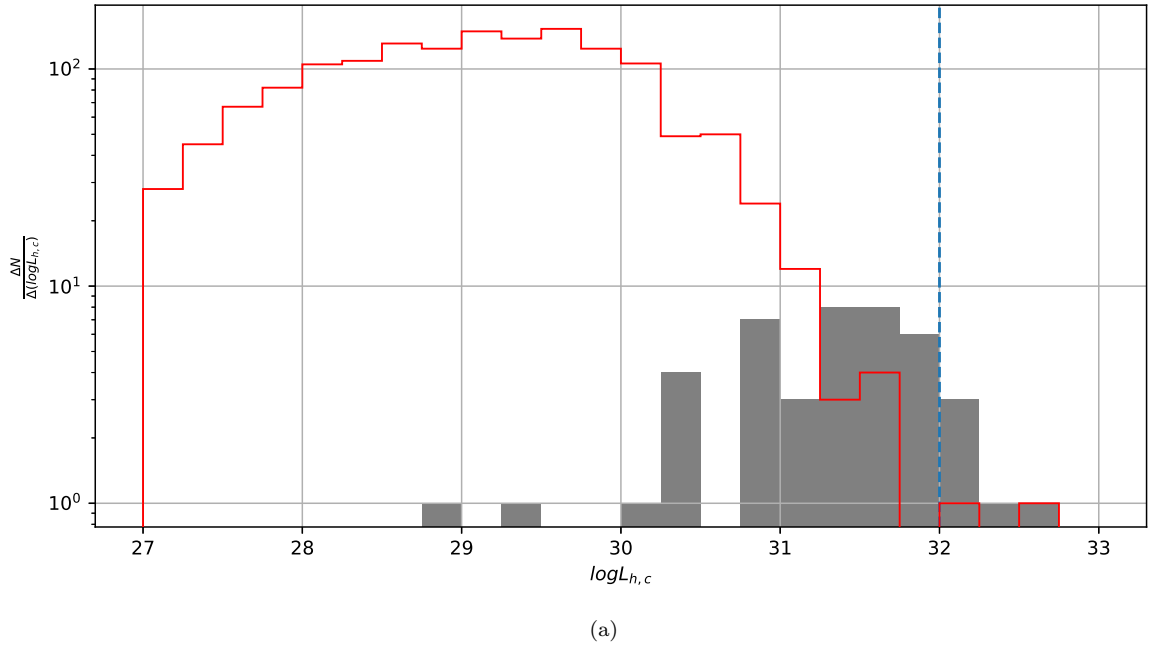}{\textwidth}{(a)}
    \fig{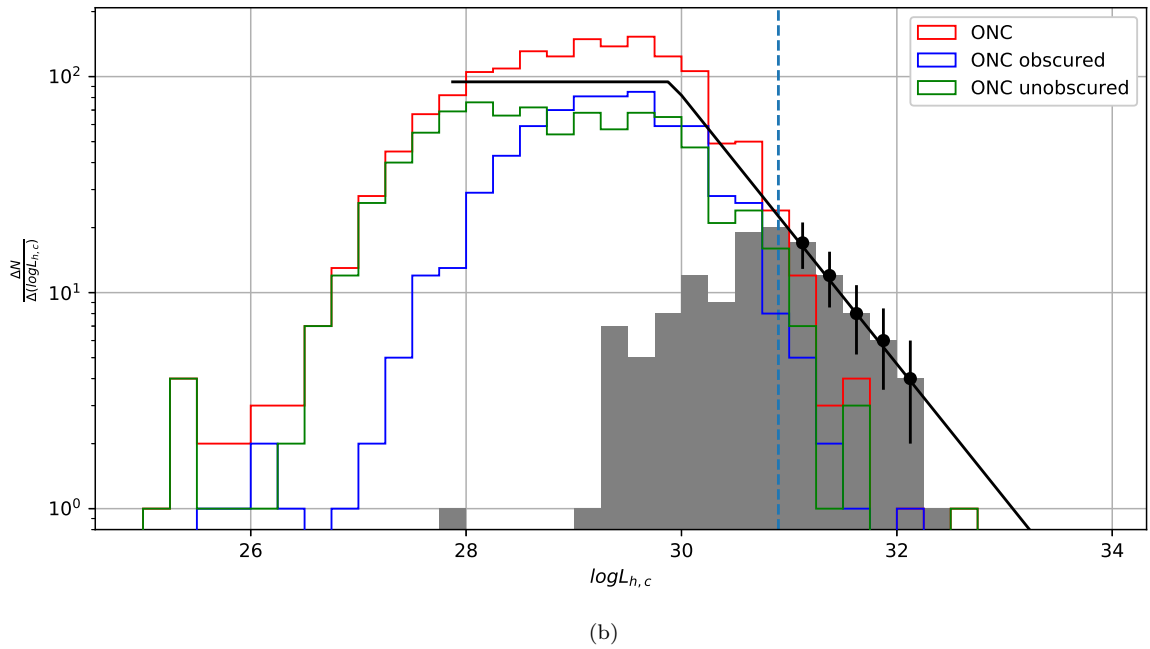}{\textwidth}{(b)}
\caption{
The X-ray luminosity functions of (a) G34.4 and (b) G35.4 (grey histograms) are compared with the well-known Orion Nebula Cluster XLF (red histogram).
Width of each luminosity bin is 0.25 dex, which means the value of XLF is also integrated by 0.25.  The black solid line in subplot (b) represents the obscured XLF of G35.4 based on the simulations, and the blue dash line shows the completeness limit of G35.4.
The completeness limit of G34.4 was estimated to be about 32.0 (blue dash line). With luminosities of only 5 sources available above this limit, estimation of its power law slope was not performed.
\label{fig:xlf}}
\end{figure}

The XLF histogram of G35.4 is shown in Figure \ref{fig:xlf}b, which is adequately sampled because of its longer exposure time compared to G34.4 (shown in Figure \ref{fig:xlf}a).  The completeness limit was estimated to be ${\rm log}\: L_{h,c}\sim30.9$ after a Monte Carlo simulation (10,000 runs) using MARX with the assistance of the spectral file of a typical X-ray point source found in the G35.4 observation. A simulated point source was located at a random position in each simulation run.  Then the simulation was used to measure the probability of whether this point source is detected by the $wavdetect$ script.

To estimate the population of G35.4, we start with the estimation of
the slope of the high luminosity end of its XLF, which was assumed to
follow a power-law distribution\citep{Wang11}.  Parameters of the power-law distribution were calculated by directly fitting the luminosity values to the model
described by \citet{mas09}, which can be simply explained as
obtaining a unique root of
\begin{equation}
- \frac{n}{1-\hat\alpha_{ML}}+n \frac{Z^{1-\hat\alpha_{ML}}\log Z - Y^{1-\hat \alpha_{ML}}\log Y}{Z^{1- \hat \alpha_{ML}}-Y^{1-\hat\alpha_{ML}}}-T = 0,
\end{equation}
where $X_n$ is the data set, and $Y = {\rm min}(X_i)$, $Z = {\rm max}(X_i)$, and
$T=\sum^n_{i=1}\log X_i$.

The result of this XLF comparison is shown in Figure \ref{fig:xlf}.  
The median of the power law slope $\Gamma$ was $-0.62\pm0.24$ (black solid line in Figure \ref{fig:xlf}b). 
For comparison, the COUP XLF with luminosity evaluated in the same energy band shows $\Gamma\sim -1.4$ in the luminosity range $30.5\lesssim {\rm log} \ L_{h,c} \lesssim 31.25$.  We also show the XLFs for the unobscured (or lightly obscured) ONC population (green histogram in Figure \ref{fig:xlf}b) with $\Gamma\sim -1.0$ and the heavily obscured population (blue histogram in Figure \ref{fig:xlf}b) with $\Gamma\sim -2.0$.

There appears to be a flatter XLF slope, which implies there are more X-ray luminous sources than expected from the XLF of the ONC cluster, or equivalently, a top heavy IMF for IRDC G35.4.  We estimate its IMF using the dereddened mass on the isochrone in the color-magnitude diagrams for G35.4 (Figure \ref{fig:twomass}d), which implies that the high-mass end IMF slope is $\sim$-0.6, compared to a value of -1.2 quoted for the ONC \citep{muench02}. This is consistent with the excess of XLF at the luminous end. However, as we cautioned earlier, such mass estimates heavily rely on the adopted distance, age and the amount of extinction.

There are several uncertainties in deriving the XLF that worth noting and caution us from over-interpreting the deviation from XLF. First of all, the contaminants such as AGNs in the X-ray detections have hard spectra and may have significant flux in the hard band compared to the young stars. Given the limited sensitivity of 2MASS and the number of NIR counterparts, we choose not to exclude sources without NIR counterparts in the XLF. Second, the flaring sources have more weight on the hard band luminosity in our relatively short observations while the impact of brief flares during the long exposure for COUP XLF is much less. A third uncertainty comes from the distance, where the luminosity of foreground sources could be overestimated if adopted larger distances, and consequently lead to excess in the high luminosity end of the XLF. Lastly, \citet{feig05} noted that the obscured ONC population shows a deficit in lower luminosity stars, implying stars with localized circumstellar column densities reaching $10^{24}$ cm$^{-2}$ could be missing even with the depth of COUP in the hard band. For YSOs deeply embedded in the IRDC, similar YSOs with the highest $N_H$ could well remain undetected in our observation.

Considering that the stellar population of the ONC within a radius of 2~pc, i.e., the region probed by the COUP XLF, has a mass of $2,400\:M_\odot$ \citep{dario2014}, i.e., with about 4,600 stars (including unresolved binaries), assuming a \citet{kroupa2011}
IMF from 0.01 to 100~$M_\odot$ and a binary fraction of 50\%, then the
stellar population of G35.4 (within the region probed by the Chandra
FOV, i.e., about 14~pc on a side) is estimated to have a current mass
of 1,700~$M_\odot$ and $\sim3,300$ members.  However, note that this estimate is an upper limit, given the
possibility of contamination of the XLF of G35.4 with X-ray sources that are not YSOs.

The total mass of the cloud IRDC G35.4 from MIR and NIR extinction mapping is 16,700~$M_\odot$ \citep{kainulainen}. The mass in the larger-scale region probed by the
\chandra{} FOV is somewhat larger, $\sim33,900\:M_\odot$. Thus the estimated star formation efficiency of the region from the XLF is $\epsilon_*\leq 5\%$. This estimate is consistent with the view that
the star (cluster) formation is in a relatively early stage in this IRDC.  For example, in the ONC the total stellar mass is estimated to be $\sim3,000\:M_\odot$ inside a 3~pc radius, along with a similar gas mass \citep{dario2014}, yielding $\epsilon_*\sim 50\%$, i.e., the ONC is at a much later evolutionary stage of star cluster formation.

In the context of other IRDCs, 
an estimate of the YSO content of the more distant (5~kpc) IRDC G028.37+00.07 (cloud C from the \citet[][]{butler2009} sample) has been made recently by \citet{moser20}, who used $70\:{\rm \mu m}$ identified point sources to characterize the protostellar population. They
estimated that this currently-observed protostellar population will
achieve a total stellar mass of between about 1\% and 3\% of the current IRDC mass of $70,000\:M_\odot$, which is similar to the limits we have derived for Cloud H from its XLF.

\section{Conclusions}

We present a high spatial resolution X-ray study of two IRDCs, namely G34.4 and G35.4 using \chandra{} observations. Our main findings are as
follows:

1.  We detect 112 valid X-ray point sources towards G34.4 and 209 towards G35.4.  Through positional
coincidence matching with 2MASS, WISE and GLIMPSE catalogs, we find 53\% and 59\% of these sources in G34.4 and in G35.4 have corresponding infrared counterparts, respectively.
No centrally concentrated clusters were revealed, unlike the young massive star clusters identified in other surveys of star forming regions (e.g., COUP, MYStIX).  Nevertheless the X-ray sources are associated with the dark filament and distributed in the surroundings.  This is consistent with the expectation that IRDCs are associated with early evolutionary stage of cluster formation. 

2. The location of IR counterparts in the color-magnitude diagram indicates a population of obscured 1--2 Myr old PMS stars, which tend to be of intermediate mass and high mass for both IRDCs.
Two Class $\uppercase\expandafter{\romannumeral2}$ counterparts to X-ray sources were identified in G34.4 and ten Class $\uppercase\expandafter{\romannumeral2}$ sources in G35.4.  For G35.4, most of them are located in or near dark filaments in the \spitzer{} image. They also show significant K-band excess in the color-color diagram.

3. The derived hard-band corrected XLF for G35.4 is compared with the
XLFs of the ONC. The underlying X-ray emitting population associated
with G35.4 is estimated to be at most 70\% of that of the ONC, using the COUP XLF as a calibrator. 
This is an upper limit due to potentially greater contamination of the IRDC XLF with non-YSO sources.  
These results imply that a substantial population of YSOs
could be present in the $\sim14$~pc-scale region around the IRDC, with
a mass of up to $\sim1,700\:M_\odot$. However, this is still a small
fraction, $\lesssim 5\%$ of the gas mass in the region, again
indicating the cloud is in an early stage of its star formation.

We thank the anonymous referee for providing helpful comments that greatly improve the clarity of our manuscript. JW acknowledges support by the National Key R\&D Program of China (2016YFA0400702) and the National Science Foundation of China (NSFC) grants U1831205, 12033004. JCT acknowledges support from Chandra/NASA grants GO3-14009A and GO7-18009A, NSF grant AST-2009674 and ERC Advanced Grant 788829 (MSTAR). This research has made use of data obtained from the Chandra Data Archive, and software provided by the Chandra X-ray Center (CXC) in the application packages CIAO. This work is based in part on GLIMPSE survey made with the {\em Spitzer} Space Telescope, which is operated by the Jet Propulsion Laboratory, California Institute of Technology under a contract with NASA. This work also makes use of data products from the Wide-field Infrared Survey Explorer, which is a joint project of the University of California, Los Angeles, and the Jet Propulsion Laboratory/California Institute of Technology, funded by NASA. 

\facilities{CXO (ACIS), Spitzer (IRAC)}
\software{CIAO \citep[v4.11; ][]{ciao}, XSPEC \citep[v12.10.1f; ][]{Arnaud96}, ACIS Extract \& TARA \citep{broos02}, astropy \citep{astropy}}

\begin{deluxetable*}{ccccccccccccc}
\tablecaption{X-Ray Spectroscopy for Selected Sources of G34.4: Thermal Plasma Fits \label{tab:g344fit}}
\tablewidth{700pt}
\tablenum{5}
\tabletypesize{\tiny}
\tablehead{\\
 \multicolumn{4}{c}{\multirow{2}{*}{Source}}&\multicolumn{3}{c}{\multirow{2}{*}{Spectral Fit}}&\multicolumn{5}{c}{\multirow{2}{*}{X-ray Luminosities}}&\multirow{4}{*}{Note}\\
    \cmidrule(r){1-4}
    \cmidrule(r){5-7}
    \cmidrule(r){8-12}
    &&&&
    \multirow{2}{*}{\shortstack{$log N_H$\\($cm^{-2}$)}}&\multirow{2}{*}{\shortstack{kT\\(keV)}}
    &\multirow{2}{*}{\shortstack{log$EM$\\($cm^{-3}$)}}&\multirow{2}{*}{\shortstack{log$L_s$}}&
    \multirow{2}{*}{\shortstack{log$L_h$}}&\multirow{2}{*}{\shortstack{log$L_{h,c}$}}&\multirow{2}{*}{\shortstack{log$L_t$}}&
    \multirow{2}{*}{\shortstack{log$L_{t,c}$}}\\
    Seq&CXOU J&Net Counts&Signif&\\
    (1)&(2)&(3)&(4)&(5)&(6)&(7)&(8)&(9)&(10)&(11)&(12)&(13)
} 
\startdata
1  &185246.09+012520.5&31.2&4.9&$21.7^{+0.3}$&$0.9_{-0.7}^{+1.3}$&$53.2_{-0.3}^{+0.3}$&30.9&30.3&30.3&31.0&31.3\\
2  &185248.28+012742.0&90.1&9.0&$20.3^{+1.3}$&$2.4_{-1.5}^{+3.8}$&$53.4_{-0.1}^{+0.2}$&31.3&31.2&31.2&31.6&31.6\\
4  &185250.09+012814.0&26.8&4.3&$22.9^{+0.4}_{-0.3}$&$9.5_{-0.2}$&$53.5_{-0.2}^{+1.8}$&29.5&31.4&31.6&31.4&31.8&Class $\uppercase\expandafter{\romannumeral2}$\\
5  &185251.54+012112.1&15.9&3.3&$22.8^{+0.3}_{-0.5}$&$10.0$&$53.2_{-0.3}^{+0.2}$&29.3&31.1&31.3&31.1&31.5\\
6  &185256.83+012430.3&24.7&4.5&$23.1^{+0.3}_{-0.2}$&$6.9_{-2.7}$&$53.6_{-0.2}^{+0.5}$&28.7&31.4&31.6&31.4&31.9\\
11 &185302.43+013153.0&12.9&2.7&$22.6^{+0.6}_{-0.4}$&$8.4_{-0.2}$&$53.0_{-0.4}^{+1.3}$&29.6&31.0&31.1&31.0&31.3\\
12 &185302.84+011750.2&31.9&5.0&$23.0^{+0.3}_{-0.2}$&$9.5_{-2.6}$&$53.7_{-0.2}^{+0.5}$&29.4&31.6&31.8&31.6&32.0\\
15 &185304.89+012621.7&12.7&3.2&$23.5_{-0.3}$&$2.8_{-1.3}$&$54.2_{-0.8}^{+0.6}$&26.9&31.4&32.0&31.4&32.3\\
16 &185304.91+013016.1&14.0&3.1&$22.4^{+0.3}_{-0.5}$&$3.5_{-1.3}$&$53.0_{-0.4}^{+0.6}$&29.9&30.8&30.9&30.9&31.2\\
18 &185308.23+012934.0&10.0&2.6&$23.5_{-0.3}$&$9.5_{-2.3}$&$53.4_{-0.3}^{+0.5}$&26.2&31.1&31.5&31.1&31.7\\
\enddata
\tablecomments{Col. (1): X-ray source number. Col. (2): IAU designation. 
Cols.(3): Estimated net counts extracted in the total energy band (0.5-8 keV).
Cols.(4): Photometric significance computed as (net counts)/(upper error on net counts).
Cols. (5)-(7): Estimated column density, plasma energy, and the plasma emission measure.
Cols. (8)-(12): Observed and corrected for absorption X-ray luminosities. 
Table 5 is available in its entirety in its machine readable format. 
A portion is shown here for guidance regarding its form and content.}
\end{deluxetable*}

\startlongtable
\begin{deluxetable*}{ccccccccccccc}
\tablecaption{X-Ray Spectroscopy for Selected Sources of G35.4: Thermal Plasma Fits \label{tab:g354fit}}
\tablewidth{700pt}
\tablenum{6}
\tabletypesize{\tiny}
\tablehead{\\
 \multicolumn{4}{c}{\multirow{2}{*}{Source}}&\multicolumn{3}{c}{\multirow{2}{*}{Spectral Fit}}&\multicolumn{5}{c}{\multirow{2}{*}{X-ray Luminosities}}&\multirow{4}{*}{Note}\\
    \cmidrule(r){1-4}
    \cmidrule(r){5-7}
    \cmidrule(r){8-12}
    &&&&
    \multirow{2}{*}{\shortstack{$log N_H$\\($cm^{-2}$)}}&\multirow{2}{*}{\shortstack{kT\\(keV)}}
    &\multirow{2}{*}{\shortstack{log$EM$\\($cm^{-3}$)}}&\multirow{2}{*}{\shortstack{log$L_s$}}&
    \multirow{2}{*}{\shortstack{log$L_h$}}&\multirow{2}{*}{\shortstack{log$L_{h,c}$}}&\multirow{2}{*}{\shortstack{log$L_t$}}&
    \multirow{2}{*}{\shortstack{log$L_{t,c}$}}\\
    Seq&CXOU J&Net Counts&Signif&\\
    (1)&(2)&(3)&(4)&(5)&(6)&(7)&(8)&(9)&(10)&(11)&(12)&(13)
} 
\startdata
1  &185629.72+020437.9&36.1&3.7&$22.1^{+0.6}$&$2.6_{-0.5}$&$52.8_{-0.4}^{+1.3}$&30.1&30.6&30.6&30.7&31.0&\\
2  &185634.65+020942.7&25.7&3.3&$23.5_{-0.9}$&$0.7_{-0.4}$&$55.2_{-9.3}^{+1.1}$&26.9&30.6&32.0&30.6&33.3&\\
3  &185635.12+020729.6&39.5&4.6&$23.5_{-0.2}$&$10.0_{-0.2}$&$53.4_{-0.2}^{+0.1}$&25.7&31.0&31.5&31.0&31.7&\\
4  &185635.12+021528.3&13.3&2.4&$23.5^{+-0.8}$&$10.0_{-0.2}$&$53.7_{-7.8}^{+0.2}$&26.0&31.4&31.8&31.4&32.0&\\
5  &185638.11+020931.5&18.1&2.7&$22.6^{+0.6}_{-0.5}$&$5.9_{-1.0}$&$52.5_{-0.4}^{+1.1}$&29.2&30.4&30.5&30.4&30.7&\\
6  &185639.39+020837.3&36.9&4.7&$21.4^{+0.8}$&$2.0_{-0.9}^{+5.1}$&$52.5_{-0.2}^{+0.4}$&30.2&30.1&30.2&30.5&30.6&\\
7  &185643.95+021331.6&17.2&2.3&$22.9^{+0.6}_{-0.4}$&$0.9_{-0.2}^{+2.2}$&$53.8_{-7.9}$&29.4&30.3&30.8&30.4&31.9&\\
8  &185643.98+020655.3&52.8&6.7&$22.6^{+0.2}_{-0.2}$&$10.0_{-4.0}$&$53.0_{-0.1}^{+0.1}$&29.6&31.0&31.1&31.0&31.3&\\
9  &185644.00+020245.2&49.2&5.6&$21.0$&$2.9_{-1.6}^{+7.6}$&$52.7_{-0.1}^{+0.1}$&30.5&30.5&30.6&30.8&30.9&\\
10 &185645.63+021105.6&17.1&2.8&$22.2^{+0.2}$&$0.8_{-0.3}^{+5.6}$&$52.8_{-0.8}^{+1.4}$&29.9&29.7&29.8&30.1&30.9&\\
\enddata
\tablecomments{Col. (1): X-ray source number. Col. (2): IAU designation. 
Cols.(3): Estimated net counts extracted in the total energy band (0.5-8 keV).
Cols.(4): Photometric significance computed as (net counts)/(upper error on net counts).
Cols. (5)-(7): Estimated column density, plasma energy, and the plasma emission measure.
Cols. (8)-(12): Observed and corrected for absorption X-ray luminosities. 
Table 6 is available in its entirety in its machine readable format. 
A portion is shown here for guidance regarding its form and content.}
\end{deluxetable*}

\begin{longrotatetable}
\begin{deluxetable*}{ccccccccccccccccc}
\tablecaption{Stellar Counterparts of G34.4\label{tab:g344sc}}
\tablewidth{700pt}
\tablenum{7}
\tabletypesize{\tiny}
\tablehead{\\
\multicolumn{2}{c}{\multirow{2}{*}{Source}}&\multicolumn{14}{c}{INFRARED PHOTOMETRY}\\
\cmidrule{3-16}
&&\multicolumn{4}{c}{2MASS}&\multicolumn{5}{c}{WISE}&\multicolumn{5}{c}{GLIMPSE}&\multirow{3}{*}{NOTE}\\
\cmidrule(r){1-2}
\cmidrule(r){3-6}
\cmidrule(r){7-11}
\cmidrule(r){12-16}
\multirow{2}{*}{Label}&\multirow{2}{*}{CXOU J}&
\multirow{2}{*}{2MASS}&\multirow{2}{*}{\shortstack{J\\(mag)}}&\multirow{2}{*}{\shortstack{H\\(mag)}}&\multirow{2}{*}{\shortstack{K\\(mag)}}&
\multirow{2}{*}{WISE}&\multirow{2}{*}{\shortstack{W1\\(mag)}}&\multirow{2}{*}{\shortstack{W2\\(mag)}}&\multirow{2}{*}{\shortstack{W3\\(mag)}}&\multirow{2}{*}{\shortstack{W4\\(mag)}}&
\multirow{2}{*}{GLIMPSE}&\multirow{2}{*}{\shortstack{3.6\\(mag)}}&\multirow{2}{*}{\shortstack{4.5\\(mag)}}&\multirow{2}{*}{\shortstack{5.8\\(mag)}}&\multirow{2}{*}{\shortstack{8.0\\(mag)}}\\
\\(1)&(2)&(3)&(4)&(5)&(6)&(7)&(8)&(9)&(10)&(11)&(12)&(13)&(14)&(15)&(16)&(17)
} 
\startdata
1&185246.09+012520.5&18524608+0125196&11.291&11.007&10.873&J185246.12+012520.6&10.537&10.531&9.638&9.638&G034.3492+00.3525&10.785&10.687&10.700&10.828&\\
2&185248.28+012742.0&18524832+0127427&13.093&12.405&12.197&J185248.36+012742.2&12.067&11.966&10.127&10.127&G034.3888+00.3623&12.027&11.994&11.986&11.985&\\
3&185249.99+012445.6&-&-&-&-&-&-&-&-&-&-&-&-&-&-&\\
4&185250.09+012814.0&18525009+0128138&16.585&13.863&12.612&J185250.39+012813.1&10.321&10.014&10.375&10.375&G034.3999+00.3598&11.855&11.608&11.495&10.896&Class $\uppercase\expandafter{\romannumeral2}$\\
5&185251.54+012112.1&-&-&-&-&-&-&-&-&-&-&-&-&-&-&\\
6&185256.83+012430.3&-&-&-&-&-&-&-&-&-&-&-&-&-&-&\\
7&185258.22+012411.5&-&-&-&-&-&-&-&-&-&-&-&-&-&-&\\
8&185259.83+012208.2&18525982+0122074&14.815&13.585&13.159&-&-&-&-&-&G034.3277+00.2773&12.843&12.824& & &\\
9&185300.43+012145.6&18530052+0121448&16.517&15.969&14.889&-&-&-&-&-&G034.3235+00.2718&13.387&12.996& & &\\
10&185301.02+012844.5&-&-&-&-&J185300.92+012848.8&9.585&9.230&8.923&8.923&G034.4285+00.3229&13.738&12.633& & &\\
\enddata
\tablecomments{Cols. (1)-(2) reproduce the sequence number and source identification from Tables 3 and 5. 
Cols. (3)-(6) provide NIR identifications and JHK photometry from 2MASS. 
Cols. (7)-(11) provide WISE source names and photometry in 4 channels.
Cols. (17) provide hard ratio for source without stellar counterparts and marks for Class $\uppercase\expandafter{\romannumeral2}$.
Cols. (12)-(16) provide MIR identifications and photometry in 4 channels of \spitzer{}/IRAC from GLIMPSE.
Table 7 is available in its entirety in its machine readable format. 
A portion is shown here for guidance regarding its form and content.}
\end{deluxetable*}
\end{longrotatetable}

\begin{longrotatetable}
\begin{deluxetable*}{ccccccccccccccccc}
\tablecaption{Stellar Counterparts of G35.4\label{tab:g354sc}}
\tablewidth{700pt}
\tablenum{8}
\tabletypesize{\tiny}
\tablehead{
\multicolumn{2}{c}{\multirow{2}{*}{Source}}&\multicolumn{14}{c}{INFRARED PHOTOMETRY}\\
\cmidrule{3-16}
&&\multicolumn{4}{c}{2MASS}&\multicolumn{5}{c}{WISE}&\multicolumn{5}{c}{GLIMPSE}&\multirow{3}{*}{NOTE}\\
\cmidrule(r){1-2}
\cmidrule(r){3-6}
\cmidrule(r){7-11}
\cmidrule(r){12-16}
\multirow{2}{*}{Label}&\multirow{2}{*}{CXOU J}&
\multirow{2}{*}{2MASS}&\multirow{2}{*}{\shortstack{J\\(mag)}}&\multirow{2}{*}{\shortstack{H\\(mag)}}&\multirow{2}{*}{\shortstack{K\\(mag)}}&
\multirow{2}{*}{WISE}&\multirow{2}{*}{\shortstack{W1\\(mag)}}&\multirow{2}{*}{\shortstack{W2\\(mag)}}&\multirow{2}{*}{\shortstack{W3\\(mag)}}&\multirow{2}{*}{\shortstack{W4\\(mag)}}&
\multirow{2}{*}{GLIMPSE}&\multirow{2}{*}{\shortstack{3.6\\(mag)}}&\multirow{2}{*}{\shortstack{4.5\\(mag)}}&\multirow{2}{*}{\shortstack{5.8\\(mag)}}&\multirow{2}{*}{\shortstack{8.0\\(mag)}}\\\\
(1)&(2)&(3)&(4)&(5)&(6)&(7)&(8)&(9)&(10)&(11)&(12)&(13)&(14)&(15)&(16)&(17)
} 
\startdata
1&185629.72+020437.9&18562963+0204392&12.570&13.257&12.483&-&-&-&-&-&-&-&-&-&-&\\
2&185634.65+020942.7&-&-&-&-&J185634.75+020941.7&10.391&9.828&9.595&9.595&G035.4416-00.1575&9.899&9.735&9.179&9.200&\\
3&185635.12+020729.6&-&-&-&-&-&-&-&-&-&-&-&-&-&-&\\
4&185635.12+021528.3&-&-&-&-&J185635.08+021523.9&10.920&10.938&8.371&8.371&-&-&-&-&-&\\
5&185638.11+020931.5&-&-&-&-&-&-&-&-&-&G035.4458-00.1710&13.268&13.118&12.589& &\\
6&185639.39+020837.3&18563939+0208365&14.837&13.207&12.534&J185639.40+020836.5&12.229&12.380&10.043&10.043&G035.4344-00.1830&11.930&11.836&11.655&12.082&\\
7&185643.95+021331.6&-&-&-&-&-&-&-&-&-&-&-&-&-&-&\\
8&185643.98+020655.3&-&-&-&-&-&-&-&-&-&-&-&-&-&-&\\
9&185644.00+020245.2&18564389+0202453&14.421&13.582&13.278&-&-&-&-&-&G035.3561-00.2442&12.784&12.827& & &\\
10&185645.63+021105.6&18564573+0211058&16.510&14.969&14.247&J185645.36+021106.3&11.689&11.593&10.412&10.412&-&-&-&-&-&\\
\enddata
\tablecomments{Cols. (1)-(2) reproduce the sequence number and source identification from Tables 4 and 6. 
Cols. (3)-(6) provide NIR identifications and JHK photometry from 2MASS. 
Cols. (7)-(11) provide WISE source names and photometry in 4 channels.
Cols. (12)-(16) provide MIR identifications and photometry in 4 channels of \spitzer{}/IRAC from GLIMPSE.
Cols. (17) provide hard ratio for source without stellar counterparts and marks for Class $\uppercase\expandafter{\romannumeral2}$.
Table 8 is available in its entirety in its machine readable format. 
A portion is shown here for guidance regarding its form and content.}
\end{deluxetable*}
\end{longrotatetable}
\clearpage

\end{document}